\documentclass[article,11pt, reprint, superscriptaddress, nofootinbib, amsmath,amssymb, prd,onecolumn]{revtex4-2}

\pdfoutput=1
\usepackage{graphicx}
\usepackage{amssymb}
\usepackage{amsmath}
\usepackage{mathtools}
\usepackage{breqn}
\usepackage{color,soul}
\usepackage{bm}
\usepackage{latexsym}
\usepackage{float}
\usepackage{textcomp}
\usepackage{hyperref}
\usepackage{color}
\usepackage{float}
\usepackage{blindtext}
\usepackage{enumitem}
\usepackage{xcolor}
\usepackage[section]{placeins}
\usepackage{float}
\usepackage{siunitx}
\usepackage{tikz} 
\usepackage{pgfplots}
\usepackage{graphicx}
\usepackage{cases}
\usepackage{comment}
\usepackage{multirow}
\usepackage{parskip}
\pgfplotsset{compat=newest} 
\usepgfplotslibrary{units} 

\usepackage{geometry}
\begingroup\makeatletter
\catcode`,=\active
\global\let\breqn@comma,
\protected\gdef,{\ifmmode\expandafter\breqn@comma\else\expandafter\active@comma\fi}
\endgroup

\usepackage{calligra}
\DeclareMathAlphabet{\mathcalligra}{T1}{calligra}{m}{n}
\DeclareFontShape{T1}{calligra}{m}{n}{<->s*[2.2]callig15}{}

\usepackage[mathscr]{euscript}
\usepackage[export]{adjustbox}
\usepackage{soul}

\newcommand{\beq}{\begin{equation}}
\newcommand{\eeq}{\end{equation}}
\newcommand{\bea}{\begin{eqnarray}}
\newcommand{\eea}{\end{eqnarray}}
\newcommand{\yt}[1]{\textcolor{blue}{#1}}

\def\bed{\begin{dmath}} 
\def\eed{\end{dmath}}

\def\l{\left}
\def\r{\right}
\def\d{{\mathrm{d}}}

\def\Mpl{M_{_{\mathrm{Pl}}}}

\begin{document}

\title{Understanding large scale CMB anomalies with the generalized non-minimal derivative coupling during inflation}
\author{Yashi Tiwari}
\email{yashitiwari@iisc.ac.in}
\affiliation{Joint Astronomy Programme, Department of Physics, Indian Institute of Science, C. V. Raman Road, Bangalore 560012, India}
\author{Nilanjandev Bhaumik}
\email{nilanjandev@iisc.ac.in}
\affiliation{Department of Physics, Indian Institute of Science, C. V. Raman Road, Bangalore 560012, India}
\author{Rajeev Kumar Jain}
\email{rkjain@iisc.ac.in}
\affiliation{Department of Physics, Indian Institute of Science, C. V. Raman Road, Bangalore 560012, India}

\maketitle


We study the observational implications of a class of inflationary models wherein the inflaton is coupled to the Einstein tensor through a generalized non-minimal derivative coupling (GNMDC). 
In particular, we explore whether these models can generate suitable features in the primordial spectrum of curvature perturbations as a possible explanation for the large-scale anomalies associated with the angular power spectrum of CMB temperature anisotropies.
We derive model-independent constraints on the GNMDC function for such a scenario, considering both the scalar and tensor perturbations.
We modify CosmoMC to accommodate our GNMDC framework and  investigate different  classes of inflationary models using a fully consistent numerical approach. We find that the hilltop-quartic model with a specific choice of the GNMDC function provides a considerable improvement over the best-fit reference $\Lambda$CDM model with a nearly scale-invariant power spectrum. 
While the large-scale structure observations should be able to provide independent constraints, future CMB experiments, such as CMB-S4 and CMB-Bharat, are expected to  constrain further the parameter space of such beyond canonical single-field inflationary models.



\section{Introduction}
\label{intro}

Our current understanding of the universe on large scales is remarkably well described by the standard model of cosmology -- a relatively simple six-parameter $\Lambda$CDM model, supplemented by an inflationary epoch in the very early universe \cite{Starobinsky:1980te, Guth:1980zm, Linde:1981mu, Albrecht:1982wi, Linde:1983gd}. Besides providing an elegant explanation to the numerous shortcomings of the hot Big Bang theory, an epoch of inflation also provides a causal and efficient mechanism for the origin of primordial density perturbations which source the anisotropies in the Cosmic Microwave Background (CMB) radiation and later act as seeds for the formation of the large scale structures in the universe \cite{Starobinsky:1979ty, Mukhanov:1981xt, Tegmark:2004qd, Bassett:2005xm, Baumann:2009ds, Sriramkumar:2009kg, Martin:2013tda}.  For these reasons, the inflationary paradigm is now widely considered a crucial part of the concordance model of cosmology. The simplest inflationary models are based on a minimally coupled slowly rolling scalar field, called the {\it inflaton}, with a slowly varying potential.
Such slow-roll inflationary models predict a nearly scale-invariant power spectrum of primordial curvature perturbations with very small non-Gaussianities that are in excellent  agreement with the ever-increasing precision measurements of the CMB anisotropies by WMAP and Planck \cite{WMAP:2012nax, Planck:2018jri, Planck:2019kim, BICEP:2021xfz}.

In the CMB measurements at large angular scales, several unexpected features, collectively known as CMB anomalies, have also been observed by COBE, WMAP and Planck \cite{WMAP:2008ttx, Schwarz:2015cma, Planck:2019evm}. Among these anomalies, one of the most notable ones is the low value of the temperature angular power spectrum at the quadrupole moment $\ell=2$, even below the cosmic variance of the $\Lambda$CDM model with a nearly scale-invariant primordial spectrum. Besides, some other outliers (localized features) also exist around $\ell \sim 20 - 30$. Some localized features also appear at smaller scales, most significantly around $\ell \sim 750$ in the TT and TE spectra, observed by Planck. Although all these features have marginal statistical significance from the Planck data, they have still generated enormous interest in the literature to understand if they could have a primordial origin in the early universe \cite{Chluba:2015bqa, Slosar:2019gvt}. Such features are strongly scale-dependent deviations from an otherwise nearly scale-invariant spectrum. Within the minimal class of canonical single field inflationary models, one of the simplest possibilities to generate these features is to briefly modify the slow roll dynamics of the inflaton, either by introducing a step in the inflaton potential \cite{Adams:2001vc, Hunt:2004vt, Hunt:2007dn, Covi:2006ci, Hamann:2007pa, Hazra:2010ve, Ashoorioon:2014yua, Hunt:2015iua, Dalianis:2021iig}, allowing an inflection point in the potential \cite{Jain:2007au, Jain:2008dw, Jain:2009pm, Qureshi:2016pjy} or imposing kinetic/fast roll initial conditions for the dynamical evolution of the inflaton field \cite{Contaldi:2003zv, Sriramkumar:2004pj, Nicholson:2007by, Hergt:2018crm, Ragavendra:2020old}. All these approaches give rise to localized features in the primordial spectrum, which usually improve the best fit over the concordance model and have been discussed extensively in the literature. Moreover, the possible origin of these primordial features has also been studied in the context of multi-field inflationary models \cite{Wands:2007bd, Achucarro:2010da}. In particular, a specific class of two field models that allow a rapid turn in the field space usually produce such features due to transient deviations from the slow roll conditions \cite{Gao:2013ota, Noumi:2013cfa, Braglia:2020fms}.

To extend beyond the minimal setup, an interesting possibility is also to use the scalar-tensor theories such as the Horndeski theory \cite{Horndeski:1974wa}. An interesting aspect of these theories is that despite higher-order terms in the Lagrangian, they yield second-order field equations and, thus, remain free of the Ostrogradsky instability. Within the framework of Horndeski theory, in this paper, we explore specific inflationary scenarios involving coupling of the derivative of the inflaton field to the Einstein tensor with a generalized coupling function, popularly known as the generalized non-minimal derivative coupling (GNMDC) \cite{Deffayet:2009wt, Deffayet:2009mn, Kobayashi:2011nu, Karydas:2021wmx}. These interactions induce gravitationally enhanced friction, affecting the inflationary dynamics of the scalar field and lead to very interesting phenomenological implications. 

The simplest case of GNMDC is called the non-minimal derivative coupling (NMDC), wherein the coupling function is a constant. It has been extensively explored, leading to interesting implications in early universe cosmology  \cite{Amendola:1993uh, Tumurtushaa:2019bmc} and black hole physics \cite{Rinaldi:2012vy, Anabalon:2013oea}. The drawback of NMDC is due to its constant presence also at the end of inflation and reheating era, which leads to dynamical instability during the reheating stage \cite{Ema:2015oaa}, affects the process of particle production and thus, changes the predictions of the standard reheating scenario \cite{Ghalee:2013ada}. Therefore, the simple NMDC term has been proven to be not so useful for successful inflationary model building. However, in a more general GNMDC setup, all these issues can be cured. The freedom to choose the form of the generalized coupling function makes these classes of GNMDC models phenomenologically attractive. In general, choosing a non-constant, field-dependent GNMDC function affects the dynamics of the inflaton field, thereby leaving distinguishable imprints in the primordial power spectrum.
For instance, the GNMDC term during inflation has been recently studied for forming primordial black holes and induced gravitational waves \cite{Fu:2019ttf, Heydari:2021gea, Heydari:2021qsr}. Including GNMDC coupling or the gravitationally enhanced friction can also make a class of inflationary models consistent with the CMB data, which are otherwise ruled out. Recently, this possibility has also been explored in the context of Higgs inflation assuming slow roll approximations \cite{Karydas:2021wmx}.

In this paper, we aim to study whether GNMDC models can help make inflationary scenarios consistent with the latest CMB data and simultaneously explore the prospects of generating localized features in the scalar power spectrum to explain the origin of large-scale features in the angular power spectrum of CMB temperature anisotropies. For simplicity, we only consider the GNMDC function a function of the scalar field $\phi$ and study three different inflationary models for this purpose.
Independent of the choice of the coupling function, we find relevant constraints on this class of GNMDC models  by keeping track of the propagation speed for scalar and tensor perturbations. From simple analytical arguments, we show that the generation of non-trivial localized features in the primordial power spectrum to explain the CMB anomalies in these models, together with avoiding any gradient instabilities and unphysical solutions, is inconsistent with subluminal propagation of both scalar and tensor perturbation modes simultaneously. Only if the tensor perturbations are marginally superluminal during the inflationary phase, it opens up the possibility of generating large-scale features of sufficient amplitude in the primordial scalar power spectrum. Staying within these constraints, we find that GNMDC models' efficiency in explaining large-scale CMB anomalies crucially depends on the functional form of the GNMDC function $\theta(\phi)$.

Unlike previous literature, which relies explicitly on the slow roll approximation in GNMDC, we adopt a fully numerical approach and develop a fast parallel computing module to exactly calculate the power spectra of scalar and tensor perturbation in these models, taking into account the  transient deviations from the slow roll conditions accurately. With this code, we study the evolution of the background and linear scalar and tensor perturbations accurately. We couple this code to the publicly available Cosmological Monte-Carlo code CosmoMC\footnotemark{}\footnotetext[1]{\url{https://cosmologist.info/cosmomc/}}
\cite{Lewis:2002ah} to compare these models with the concordance $\Lambda$CDM model and arrive at the best fit parameter constraints.
We use two Planck datasets and work with relevant likelihood combinations for our model comparison. For both cases, we show that within the GNMDC setup, characterized by our specific choice of $\theta(\phi)$, we obtain significant improvement in the fit at large scales in the Planck data. As mentioned earlier, the choice of $\theta(\phi)$ is crucial to obtain specific features in the power spectrum. Here, we primarily target the large-scale anomalies around $\ell \sim 20 - 30$ in the observations of CMB temperature anisotropies by introducing a localized feature in the GNMDC coupling function. Thus, we do not expect much improvement in the fit on relatively smaller scales or in polarization.  Nevertheless, we still check the consistency of our model with the small-scale temperature and polarization anisotropy likelihoods for two different datasets to ensure that the introduction of the GNMDC feature does not affect the fit in smaller scales or for polarization likelihood. The freedom to choose $\theta(\phi)$ leaves us with much richer dynamics, and one can further aim to achieve significant improvement on smaller scales as well in CMB. Thus it will be more relevant to study these models with the advent of future observations of CMB polarization, such as from CMB-S4 experiments which should be able to constrain these features better and provide stronger bounds on the parameters space of these models.

This paper is organized as follows. In the following section, we shall briefly discuss the details of the inflationary setup, including the GNMDC term, and derive the equations of motion of the background evolution and the linear scalar and tensor perturbations. In section \ref{section3}, we discuss various issues arising in GNMDC models, such as the gradient instability associated with the  scalar perturbations and the superluminal propagation of gravitational waves, and also provide possible  resolutions to these problems. In section \ref{section4}, we discuss our choice of the GNMDC function $\theta(\phi)$ and the various inflationary models that we work with. In section \ref{data-analysis-and-methodology}, we discuss the methodology of our numerical approach that we have developed to compare these models with the data. In section \ref{results}, we present our results for the best-fit constraints on various model parameters and the best-fit CMB power spectra. Finally, in section \ref{conclusions}, we summarise our results, conclude with a discussion and present some outlook for future work in this direction. 
In appendix \ref{appendix-1}, we present the evolution equations corresponding to the background and perturbations for the GNMDC set-up, which we use in our numerical module to compute the power spectra of scalar and tensor perturbations. 

Our notations and conventions are as follows.
We shall work with natural units such that $\hbar=c=1$, and the reduced Planck mass $\Mpl=\l(8\,\pi\, G\r)^{-1/2}$.
We shall work in the spatially flat Friedmann-Lema\^itre-Robertson-Walker~(FLRW) universe described by the following 
line element
\beq
\d s^2=-\d t^2+a^2(t)\,\d {\bf x}^2
=a^2(\tau) \l(-\d \tau^2+\d {\bf x}^2\r),\label{eq:FLRW}
\eeq
where $t$ and $\tau$ denote the cosmic time and conformal time, respectively while $a$ represents the scale factor of the $3-$dimensional spatial hypersurface.
An overdot and overprime will denote differentiation with respect to $t$ and $\tau$ coordinates.

\section{Generalized non-minimal derivative coupling (GNMDC) during inflation}
\label{horndeski}

As mentioned in the introduction, we explore the implications of a derivative coupling which can be motivated by the well-studied scalar-tensor theory like the Horndeski theory \cite{Kobayashi:2011nu, Kobayashi:2019hrl}.  The complete action for the Horndeski theory (or equivalently, for the generalized Galileons), constructed out of the metric tensor and a scalar field, is given as \cite{Kobayashi:2011nu} 
\beq
\mathcal{S} = \int d^4x \sqrt{-g}\, \sum_{i=2}^5 \mathcal{L}_{i}\,,
\eeq
where
\bea
     && \mathcal{L}_2\ =\  G_2(\phi, X),\nonumber\\
     && \mathcal{L}_3\ =\  -G_3(\phi, X)\Box{\phi},\nonumber\\
     && \mathcal{L}_4\ =\  G_4(\phi, X)R + G_{4X}(\phi, X)\Big[(\Box{\phi})^2 - (\nabla_{\mu}\nabla_{\nu}\phi)^2\Big],\nonumber\\
     && \mathcal{L}_5\ =\  G_5(\phi,X)G_{\mu\nu}\nabla^{\mu}\nabla^{\nu}\phi - \frac{1}{6}  G_{5X}\Big[(\Box{\phi})^3-3\Box{\phi}(\nabla_{\mu}\nabla_{\nu}\phi)^2+2(\nabla_{\mu}\nabla_{\nu}\phi)^3\Big], \nonumber
\eea
where $R$ is the Ricci scalar, $G_i$ are four independent arbitrary functions of $\phi$ and $X$, and $G_{iY} =\partial{G}_i/\partial{Y}$ with $Y=\{\phi,X\}$ and $X=-\partial_{\mu}\phi \partial^{\mu}\phi/2$ . For specific choices of the functions $G_i$, one can reproduce most of the second-order scalar-tensor theory as a specific case. For instance, the non-minimal coupling to gravity can be obtained by setting $G_4=G_4(\phi)$. Moreover, the Einstein-Hilbert action is already contained in this construction and can be recovered by setting $G_4 = \Mpl^2/2$. For our case, the action comprising of a GNMDC interaction during inflation can be obtained from the Horndeski setup by choosing $G_2=X-V(\phi)$, $G_3=0$, $G_4=\Mpl^2/2$, $G_5=G_5(\phi)$, and further doing integration by parts for the term $\mathcal{L}_5$, we arrive at
\beq
\mathcal{S}=\int d^4x\sqrt{-g}\left[\frac{ \Mpl^2}{2}R-\frac{1}{2}\bigg(g^{\mu\nu}-\theta(\phi)G^{\mu\nu}\bigg)\partial_{\mu}\phi\partial_{\nu}\phi -V(\phi)\right],
\label{e:action}
\eeq
where $V(\phi)$ is the potential of the inflaton field and $\theta(\phi)=-2G_{5\phi}$. 
As mentioned earlier, the case where $\theta(\phi)$ is a constant is referred to as the NMDC. 
As we shall discuss later, for a given potential, an appropriate choice of the coupling function $\theta(\phi)$ can generate interesting features in the primordial spectrum of curvature perturbations and, therefore, in the CMB angular power spectrum which can possibly explain the observed large scale features in the CMB temperature anisotropies. 

For the homogeneous, isotropic, and spatially flat FLRW metric, the background Friedmann equations can be written as \cite{Fu:2019ttf}
 \bea
 \label{frw1}
 3H^2&=&\kappa^2\l[V(\phi)+\frac{1}{2}\dot{\phi}^2\l(1+9\theta(\phi)H^2\r)\r], \\
 \label{frw2}
 -2\Dot{H}&=&\kappa^2\l[\l(1+3\kappa^2\theta(\phi)H^2-\kappa^2\theta(\phi)\dot{H}\r)\dot{\phi}^2-\kappa^2\theta'(\phi)H\dot{\phi}^3-2\kappa^2\theta(\phi)H\dot{\phi}\ddot{\phi}\r],
 \eea
where $\theta' = d\theta/d\phi$, $\kappa^2=1/\Mpl^2$ and from now on, we set $\kappa^2=1$.
The Klein-Gordon equation for $\phi$ can be obtained as
\beq
\Ddot{\phi}\l(1+3\theta(\phi)H^2\r)+3H\Dot{\phi}\l(1+\theta(\phi)(3H^2+2\Dot{H})\r)+\frac{3}{2}\theta'(\phi)\Dot{\phi}^2H^2+V'(\phi)=0,
\eeq
where $V'=dV/{d\phi}$. Evidently, all these equations reduce to the case of a single minimally coupled canonical scalar field in the absence of the GNMDC term. Moreover, the presence of the GNMDC term allows for an extra friction term in the equations of motion thereby changing the inflationary dynamics drastically. 
Using equation (\ref{frw1}), we find that the velocity of the inflaton field can be obtained  as
\beq
\label{dotphi}
\Dot{\phi}=\pm\frac{\sqrt{2\l[3H^2-V(\phi)\r]}}{\sqrt{1+9\theta(\phi)H^2}},
\eeq
which implies that the real solutions of $\Dot{\phi}$ would demand
\bea
1+9\theta(\phi)H^2 &> &0, \label{constrain1}  \\
3H^2 &>&V(\phi).
 \eea
These conditions together impose a constraint on the GNMDC function $\theta(\phi)$ in terms of the potential of the inflaton field. Therefore, in order to avoid unreal values of the inflaton velocity, we obtain the following condition that we always impose
\beq
1+3\theta(\phi)V(\phi)>0.
\label{constrain}
\eeq
Further, for a given potential $V(\phi)$ and the GNDMC function $\theta(\phi)$, we solve the background equations numerically with the number of e-folds $N$ as the time variable by using appropriate initial conditions on $\phi$, $\Dot{\phi}$ and $H$ by requiring that
\begin{itemize}
\item 
Inflation lasts for a long enough duration which determines the initial value of the inflaton field i.e. $\phi_i = \phi(N=N_i)$.
\item  
In order to have negligible deviations from slow roll dynamics at an initial time $N_i$, we set the initial condition on the Hubble parameter using equation (\ref{frw1}) as
\beq
H=\sqrt{\frac{V(\phi)}{3}+\xi}\; , \qquad 0<\xi \ll 1.
\eeq
\item 
Finally, the initial condition on $\Dot{\phi}$ is given by equation (\ref{dotphi}). 
\end{itemize}

In order to understand the background evolution for a given $V(\phi)$ and $\theta(\phi)$, we have developed a numerical module based on an exact integration of the Friedmann equations which are subject to initial conditions as mentioned above. This is required as we are interested in generating primordial features in the power spectrum of curvature perturbations that may arise due to transient deviation from slow-roll conditions. Since the  slow roll approximation is usually not able to correctly capture these features, we shall not discuss the slow roll approximation for the background quantities and the slow roll results for the power spectrum of cosmological perturbations. 

We shall now discuss the evolution of linear cosmological perturbations in this scenario and focus only on the scalar and tensor perturbations as vector perturbations usually decay during inflation unless being sourced. Following the standard cosmological perturbation theory at the linear order, the second order action for the scalar curvature perturbations $\mathcal{R}$ can be calculated as \cite{Kobayashi:2011nu, Kobayashi:2019hrl} 
\beq
\mathcal{S}^{(2)}_{{\cal R}}=\int dt\, d^3x\, a^3{\cal G}_{s}\l[\Dot{\mathcal{R}}^2-\frac{c_{s}^2}{a^2}(\partial \mathcal{R})^2\r],
\label{scalar-action}
\eeq
where $c_{s}$ is the propagation speed of the scalar modes, given by
\beq
c_{s}^2 =\frac{{\cal F}_{s}}{{\cal G}_s},
\label{speedofS}
\eeq
and ${\cal F}_s$ and ${\cal G}_s$ are defined as
 \begin{align}
{\cal F}_s&=\frac{1}{a}\frac{d}{dt}\l(\frac{a}{\Theta}{\cal G}_{_T}^2\r)-{\cal F}_{_T}, \\
{\cal G}_s&=\frac{\Sigma}{\Theta^2}{\cal G}_{_T}^2+3{\cal G}_{_T},
\end{align}
with
\begin{align}
\Sigma&=
X-3H^2+18H^2X \theta(\phi)\\
\Theta&=
H\l[1-3X \theta(\phi)\r]\\
{\cal F}_{_T}&=
1+X \theta(\phi)\\
{\cal G}_{_T}&=
1-X \theta(\phi)
\end{align}

In order to avoid the ghost and gradient instabilities associated with the curvature perturbations, we must require that ${\cal G}_{s} >0$ and $c_{s}^2 >0$ which, in turn, imposes constraints on the GNMDC coupling function $\theta(\phi)$.
In terms of the conformal time $d\tau=dt/a$, the second order action in eq. (\ref{scalar-action}) can be written as 
\beq
\mathcal{S}^{(2)}_{{\cal R}}= \int d\tau d^3x \, z_s^2\l(\mathcal{R'}^2-c_{s}^2(\partial\mathcal{R})^2\r),
\eeq
where $ z_s=a\sqrt{{\cal G}_s}$. Using this action, the Mukhanov-Sasaki equation for the Fourier modes of curvature perturbations $\mathcal{R}_k$ can be obtained as 
\beq
\mathcal{R}_k''+\frac{2z_s'}{z_s}\mathcal{R}_k'+c_s^2k^2\mathcal{R}_k=0.
\label{rk-eqn}
\eeq
The power spectrum of curvature perturbations is defined by
\beq
\mathcal{P}_{_{\mathcal{R}}}(k)=\frac{k^3}{2\pi^2}|\mathcal{R}_k|^2,
\eeq
which is evaluated at the end of inflation using an exact numerical integration of the Fourier modes equation and by imposing the appropriate initial conditions on the Fourier modes in the sub-horizon regime.

The linear tensor perturbations are described by the second-order action which is given by
\beq
\mathcal{S}^{(2)}_T=\frac{1}{8}\int dt\, d^3x\, a^3\l[{\cal G}_{T}\Dot{h}^2_{ij}-\frac{{\cal F}_{T}}{a^2}(\nabla h_{ij})^2\r],
\eeq
which in terms of conformal time can be written as,
\beq
{\mathcal{S}_T}^{(2)}= \int d\tau d^3 x \, {z_{T}}^2\l({{h'_{ij}}^2}-{c^2_T}(\nabla h_{ij})^2\r),
\eeq
where $z_{T} = \frac{a}{2} \l({\cal G}_{T}/2\r)^{1/2}$ and $c_{T}^2= {\cal F}_{T}/{\cal G}_{T}$.
Finally, the Fourier mode equation for tensor perturbations can be obtained by varying the above action which leads to 
\beq
h_k''+\frac{2z_{_T}'}{z_{_T}}h_k'+c_{_T}^2k^2h_k=0,
\eeq
and the power spectrum of tensor perturbations is defined by
\beq
\mathcal{P}_{h}(k)=2 \frac{k^3}{2\pi^2}|h_k|^2.
\eeq
Note that, the power spectra for both the scalar and tensor perturbations must be evaluated at the end of inflation or in the super-horizon limit $k/aH \ll 1$. 

\section{Sound speed of linear perturbations in GNMDC and CMB anomalies}
\label{section3}

In this section, we shall discuss our approach to avoid the issues associated with the propagation speed of linear perturbations and instabilities that arise in the presence of the GNMDC term. In particular, we work in the regime of parameter space such that we avoid the problems like gradient instability and superluminal propagation of the scalar  perturbations. 

\subsection{Sound speed of scalar perturbations }
\label{scalar-gradient}

The propagation speed of the scalar modes in GNMDC setup is given by equation (\ref{speedofS}). Upon changing the time variable to the efolds $N$ using $dN = H dt$ and defining $\phi_N={d\phi}/{dN}$, $\phi_{NN}={d^2\phi}/{dN^2}$, $H_N={dH}/{dN}$, $\theta_N(\phi)={d\theta(\phi)}/{dN}$,
$g \equiv -\frac{1}{2} \theta(\phi) H^2$
and 
$h \equiv  -\frac{1}{2} \theta_N(\phi) H^2$
together with some simplifications, we get
\begin{align}
c_{s}^2 & =
\Bigl( H \left(g \phi_N \left(\phi_N \left(g (18 g+8 h+1) \phi_N^2-4 g+8 h+2\right)
+16 g \phi_{NN}
\left(g \phi_N^2+1\right)\right)-6 g+1 \right)
\notag \\ & \quad
+4 g H_N \left(g \phi_N^2+1\right) \left(3 g \phi_N^2-1\right) \Bigr) 
/\Bigl(H\left(g \phi_N^2+1\right) \left( g (18 g+1) \phi_N^2-6 g+1 \right) \Bigr)
\end{align}
As we have discussed in the previous section, we are interested in generating features in the power spectrum due to the presence of the GNDMC term and thus, we assume that $g$ and $h$ need to be of ${\mathcal{O}}(1)$ or larger to get such large scale features. Moreover,  if we also restrict ourselves to slowly varying functions of 
$\theta(\phi)$, we can assume\footnote{We have also numerically checked that these assumptions hold quite well for the entire duration of inflation.} $|g|>|h|\gg H\gg |H_N|$ and $|g|>|h|\gg |\phi_N|>|\phi_{NN}|$. These imposed hierarchies are largely based on the requirement of having a stable inflationary phase and having a significant effect of the GNMDC term to generate localized features in the primordial power spectrum. Now, under these conditions and taking only the leading order contributions, it is possible to write $c_s^2$ as
\begin{align}
c_s^2 \approx 1 -\frac{16 g^2 \phi_{N}^2}{1-6 g}+\frac{16 g^2 \phi_{N} \phi_{NN}}{1-6 g}+\frac{8 g h \phi_N^2}{1-6 g}.
  \label{cscon1}
\end{align}
Since $|g| > {\cal O}(1)$ to get appropriate features in the scalar power spectra, the above expression further simplifies to 
\begin{align}
c_s^2 \approx 1+\frac{8}{3} g \phi_N^2-\frac{8}{3} g \phi_N \phi_{NN}-\frac{4}{3} h \phi_N^2  \approx 1+\frac{8}{3} g \phi_N^2\,.
    \label{cscon}
\end{align}
For a consistent theory {\it sans} instabilities, we need $0<c_s^2 \leq 1$, so from the above equation, it is evident that we need 
\begin{align}
0 \leq \theta(\phi) < \frac{3}{4\phi_N^2 H^2} \,.
\label{csbound}
\end{align}
Note that, while the first part $\theta(\phi) \geq 0$ is obtained to avoid the superluminal propagation of scalar perturbations, the
second part of the above inequality, $\theta(\phi)< \frac{3}{4 \phi_N^2 H^2}$ avoids the gradient instability, which is caused whenever the sound speed of scalar perturbations becomes imaginary. As studied in earlier works, the presence of such a gradient instability can cause an uncontrollable growth of scalar perturbation modes, thereby, rendering the theory inconsistent, even if the background evolution is stable \cite{Quiros:2017gsu, Hsu:2004vr}. It is, therefore, necessary to ensure that the squared sound speed of perturbation modes is always positive i.e. $c^2_s >0$.

\begin{figure}[!t]
\hskip -30pt
\includegraphics[width=0.78\columnwidth,height=9.0cm]{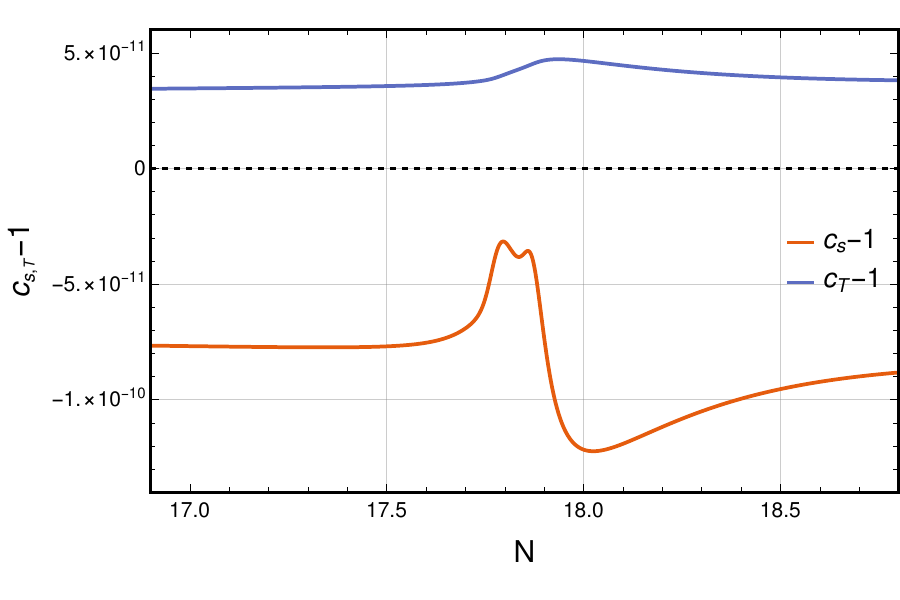}
\caption{The deviation of sound speed of scalar and tensor perturbations from unity in our GNMDC model for the best-fit parameter values of Dataset I as given in Table \ref{tab:Table4}. As is evident from the plot, the scalar modes remain subluminal throughout, while the tensor modes are slightly superluminal in our GNMDC model. The sound speeds corresponding to the best-fit values from other datasets also show similar behavior.}
\label{speed}
\end{figure}

\subsection{Sound speed of tensor perturbations}

Following the discussion of section \ref{horndeski}, the sound speed of tensor perturbations is given by
\begin{align}
c_{_T}^2& =
\frac{{\cal F}_T}{{\cal G}_T}= 
\frac{1+X \theta(\phi)}{1-X \theta(\phi)}\,.\label{speedofGW_H}
\end{align}

As long as we work with 
negative values of $\theta(\phi)$, we can infer from the above equation that  $c_{_T}^2 < 1$, but any positive $\theta(\phi)$ shall lead to superluminal GWs. The recent detection of GWs from LIGO-VIRGO collaboration and their electromagnetic counterparts lead to a very stringent bound on the sound speed of GWs \cite{LIGOScientific:2017zic, Gong:2017kim} given as,
\begin{align}
-3\times 10^{-15}<c_T-1<7\times10^{-16} \,.\label{cTbound}
\end{align}
But it is to note that these bounds are valid for the late universe wherein the observed signals are originated. One can, in principle, argue that in very early universe, these bounds can be grossly violated.

\subsection{Our adopted approach}
\label{our-approach}

While avoiding gradient instability constrains the amplitude of  $\theta(\phi)$, and therefore the amplitude of local features in the scalar power spectra, the issue of superluminality for scalar and tensor modes is rather tricky to handle. To generate appropriate wiggles in the primordial scalar power spectrum, we use a localized dip of sufficient amplitude in the function  $\theta(\phi)$. If we only consider this localized feature on top of the minimal setup, the local dip inevitably leads $\theta(\phi)$ to negative values. 
However, in the negative regime, $\theta(\phi)$ is constrained to  be $\theta(\phi) > -1/9H^2  $ from equation  (\ref{constrain1}) for avoiding unphysical solutions. On the other hand, we need the amplitude of $\theta(\phi) $ to be of ${\mathcal{O}}(1/H^2)$ to obtain any visible features in power spectrum \cite{Fu:2019ttf, Heydari:2021gea, Heydari:2021qsr}. Considering this difficulty, along with the restriction imposed by equation (\ref{csbound}) to have subluminal propagation of scalar modes, we choose to restrict ourselves to positive values of $\theta(\phi)$. One immediate setback of this approach is to have superluminal propagation of tensor modes. As is evident from equation (\ref{speedofGW_H}), it is inevitable to have superluminal tensor modes whenever we consider $\theta(\phi) \geq 0$ regime. Equation (\ref{speedofGW_H}) together with equation (\ref{csbound}) shows that in these class of models which induce non-negligible features in the scalar power spectrum, superluminalities of scalar and tensor modes are mutually exclusive. In other words, the subluminal propagation of scalar modes requires $\theta(\phi)\ge0$  which inevitably leads to superluminal GWs, and vice versa.

For the viability of our model, we restrict ourselves to the cases wherein we can completely avoid the gradient instability  and superluminality for the scalar modes but the tensor models remain superluminal.
To achieve this, we work with $\theta(\phi)>0$ and propose a possible way to obtain $\theta(\phi)>0 $ throughout the evolution by using an extra monomial term in $\theta(\phi)$ which will be present during the inflationary phase but becomes negligible towards the end of inflation. Together with this, we have a local feature term that shall be active near the horizon exit of relevant large scales. In this case, the suppression or wiggles due to the local feature term will arise with respect to the amplification coming from the monomial term and not with respect to the minimal results. With this, we can now write, 
\begin{align}
\theta(\phi)=\theta_0(\phi)+\theta_1(\phi),
\label{G5phibreak}
\end{align}
where $\theta_0(\phi)$ denotes the monomial term and $\theta_1(\phi)$ refers to the local term responsible for features in the spectrum. The advantage of using this setup is two-fold. First, the monomial term $\theta_0(\phi)$ will cancel out any negative contribution coming from $\theta_1(\phi)$, and shall ensure that $\theta(\phi)$ is always positive for consistency. 
Second, it will also control the scalar spectral index $n_s$, and the tensor-to-scalar ratio $r$. In the context of primordial black holes forming models \cite{Dalianis:2019vit} and Higgs inflation \cite{Karydas:2021wmx}, the applications of such monomial terms have been explored recently. 

In section \ref{theta-choice}, we will discuss the choice of our $\theta(\phi)$, to obtain large-scale features in the primordial power spectrum. We show that by appropriate choice of a positive $\theta(\phi)$, we always reside in the subluminal regime for scalar modes, while tensor modes remain marginally superluminal during the inflationary evolution in our GNMDC setup. Figure \ref{speed} shows the extent of violation of the superluminality of tensors in terms of $c_T-1$ for our best-fit GNMDC model, which is $\sim \mathcal{O}(10^{-11})$. Although this violates the bounds from equation (\ref{cTbound}), but as argued earlier, these bounds constrain the late universe physics, while the early inflationary phase can still have deviations from them.

\section{Choice of the GNMDC coupling function and inflationary models}
\label{section4}

\subsection{GNMDC coupling function $\theta(\phi)$}
\label{theta-choice}

\subsubsection{Choice of $\theta_0(\phi)$}
\label{g5phi0}

The monomial term $\theta_0(\phi)$ is a global term that is dynamically active throughout the duration of inflation. The motivation to use this term is to ensure that $\theta(\phi)$ remains positive as we impose this condition following the discussion of our previous section. Further, to get an unaltered reheating history and the standard radiation domination, the contribution from this term should be negligible towards the end of inflation. 
Therefore, we essentially need a  $\theta_0(\phi)$ term whose amplitude monotonically  decreases during inflation. As a simple example, we work with a monomial form of $\theta_0(\phi)$ given as,
\begin{align}
 \theta_0(\phi)=
 A_0  \left(\frac{\phi}{\phi_i}  \right)^{n}. 
\end{align}
The sign of index $n$ depends on the model for the potential of the inflaton field. For models where $\phi$ decreases with time, $n$ should be positive so that by the end of inflation, GNMDC contribution becomes insignificant. Similarly, $n$ should be negative for the potentials where $\phi$ increases with time. 
Previous studies have considered inflationary models with a monomial GNMDC term, leading to shifting in the values of $n_s$ and $r$, in comparison to their values in the minimal setup without GNMDC. In refs. \cite{Dalianis:2019vit, Karydas:2021wmx}, it has been shown that, for a given background model, such a monomial term with a larger index leads to a stronger shift in the values of $n_s$ and $r$. In this work, we fix the power law index of $\theta_0(\phi)$, with $|n|=4$, while its sign depends on the potential of the inflationary model. Later in section \ref{potential-choice}, we will explore the behavior of different inflationary potentials $V(\phi)$ in the presence of such a monomial GNMDC term to find the best one with the allowed values of $n_s$ and $r$. 
An appropriate combination of $V(\phi)$ and $\theta_0(\phi)$ leads to a nearly scale-invariant primordial power spectrum. Next, we will discuss the form of $\theta_1(\phi)$ in order to introduce desired localized features on the relevant large scales.

\subsubsection{Choice of $\theta_1(\phi)$}

The second term $\theta_1(\phi)$ is a local term whose effects would be significant only in the vicinity of the horizon exit of very large scales. While the monomial term $\theta_0(\phi)$ along with the background model of $V(\phi)$, sets the desired power law form of inflationary scalar power spectra, this local term is largely responsible for the superimposed oscillations on relevant scales. Previous studies \cite{Braglia:2021rej, Hazra:2021eqk, Braglia:2021sun} have suggested an overall suppression and superimposed oscillations at very large scales of inflationary scalar power spectra to explain the anomalies associated with the CMB temperature anisotropy power spectrum around multipoles $\ell \sim 20-30$.
Therefore, to obtain such large-scale features, we propose a suitable form of the GNMDC function $\theta_1(\phi)$, such that it has a localized dip feature with a negative amplitude. In fact, a similar  functional form of the coupling function has been used in \cite{Fu:2019ttf}, in the context of the formation of primordial black holes, where instead of a dip, a localized peak was introduced on relevant scales. In our work, we use
\begin{equation}
\theta_1(\phi)=
-\frac{A_1}{\sqrt{1+\frac{(\phi-\phi_0)^4}{\sigma}}}\,.
\label{gnmdc1} 
\end{equation}
Here, $A_1$ is an overall amplitude factor, $\phi_0$ represents the location of large-scale features, and $\sigma$ effectively determines the width of the  feature. In later sections, we shall compare this model with the data to arrive at the best-fit constraints on various parameters. 
The amplitude factor $A_1$ in eq. (\ref{gnmdc1}) can be interpreted as 
\begin{equation*}
    A_1=A_{\rm 1,max}\times f 
\end{equation*}
where $f$ is a fraction, with $ 0 \leq f \leq 1$. The quantity $A_{\rm 1,max}$ is the maximum possible amplitude factor in  $\theta_1(\phi)$, for a given $A_0$ (i.e. amplitude of $\theta_0(\phi)$) such that
\begin{equation*}
    \theta_0(\phi_{\rm min})+\theta_1(\phi_{\rm min})\ge 0\; ,
\end{equation*}
where $\phi_{\rm min}$ corresponds to the value of the scalar field at which  $\theta_1(\phi)$ attains its minimum or the most negative value. This ensures that even at the most negative values of $\theta_1(\phi)$, the overall $\theta(\phi)$ still remains positive, following the discussion of section \ref{our-approach}.  Note that, $f=0$ implies the absence of any local dip-like feature in the GNMDC function, and hence the absence of any non-trivial large-scale features in the primordial power spectra. However, any non-zero value of $f$ will induce suppression and superimposed oscillations in the power spectra on very large scales, where $f$ controls the amplitude of these features. This will be discussed at length in the next section.

\subsection{Choice of different inflationary potentials}
\label{potential-choice}

As discussed in the introduction, in order to explain the large-scale CMB anomalies, we need appropriate scalar power spectra with superimposed oscillations from a given inflationary model. Specifically, to be consistent with the CMB constraints, we need (i) an appropriate amplitude $A_s$ of the scalar power spectrum at the pivot scale \cite{Planck:2018jri}  (ii) the correct value of the scalar spectral index $n_s$ at the pivot scale \cite{Planck:2018jri} and, (iii) sufficiently low tensor to scalar ratio $r$ so as to be consistent with the observational bound \cite{Planck:2018jri, BICEP:2021xfz}. Moreover, in order to provide a better fit to the data, the power spectrum  should have suppression at large scales and superimposed oscillations at intermediate scales \cite{Covi:2006ci, Hamann:2007pa, Hazra:2014jka, Braglia:2021sun}.

\yt{}
In order to achieve these requirements, we have studied three different single-field inflationary models, in the presence of a GNMDC monomial term $\theta_0(\phi)$, to find the optimal one which satisfies all these conditions. We shall briefly discuss the advantages and disadvantages of these different models in the following subsections and finally work with the hilltop-quartic scenario for our analysis.

\begin{figure*}
\centering
\includegraphics[width=0.9\columnwidth,height=8.86cm]{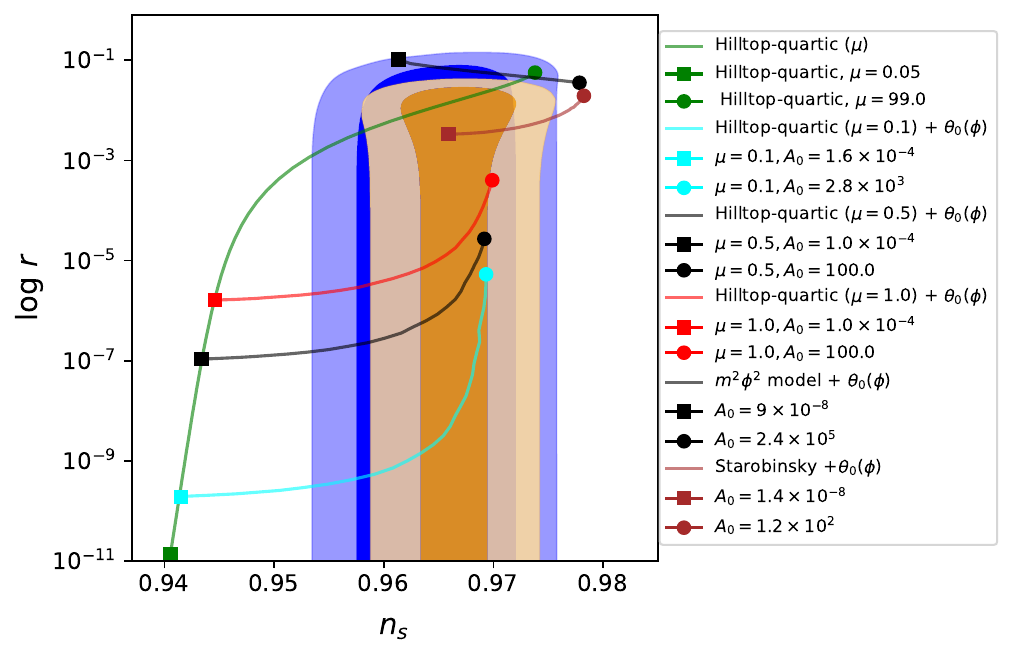}
\caption{The plot of $n_s$ vs. $r$. The blue region indicates the  1-$\sigma$ and 2-$\sigma$ contours associated with the reference $\Lambda$CDM model with Planck data only while the yellow region indicates the  1-$\sigma$ and 2-$\sigma$ contours corresponding to the Planck + BICEP recent datasets. The predictions of various inflationary models have also been displayed on top of these contours. Among these scenarios, the Straobinsky model lies at the sweet spot of the observational constraints of Planck and BICEP data. Moreover, in the presence of the monomial GNMDC term characterized by an amplitude term $A_0$, we find that both the quadratic and Starobinsky model become disfavoured with increasing $A_0$ while the hilltop-quartic model (for $\mu \lesssim 1$) provides a better fit with the data. This feature indicates that a class of models which are otherwise inconsistent with the observational constraints can possibly be made consistent with the data by introducing a monomial GNMDC term. For this reason, we finally use the hilltop-quartic model in our analysis.}
\label{ps}
\end{figure*}

\subsubsection{The quadratic model}

One of the simplest models of inflation is described by the quadratic potential
\beq
V(\phi)=\frac{1}{2} m^2 \phi^2
\eeq
This large field model leads to a large tensor to scalar ratio $r \sim 0.1 $ and has, therefore, been ruled out from recent Planck 2018 \cite{Planck:2018jri} and BICEP 2021 data release \cite{BICEP:2021xfz}.  With the addition of a monomial GNMDC term $\theta_0(\phi)$ in the Lagrangian, it is possible to decrease $r$ in this setup but that simultaneously increases the scalar spectral index $n_s$ to quite a large value, far from the best fit and thus, fails to show any improvement with the CMB data. We study the behavior of the quadratic model together with the GNMDC monomial functional form 
$\theta_0(\phi)=A_0 \phi^4$, where we vary the strength of the monomial term by varying the coefficient $A_0$. It is evident from figure \ref{ps} that even if we increase the strength of the monomial GNMDC term by increasing $A_0$, the quadratic model becomes strongly disfavoured, particularly with the recent BICEP data \cite{BICEP:2021xfz}, combined with the BAO observations. 

\subsubsection{The Starobinsky model}

The potential for the Starobinsky model, in the Einstein frame, is given by
\beq
V(\phi)=V_0 \left(1-e^{-\sqrt{2/3}\,\phi/\Mpl}\right)^2
\eeq
In the minimal setup, the Starobinsky model leads to the best-fit value of $n_s$ and a very small value of $r$, thereby sitting at the so-called sweet spot of the Planck + BICEP allowed constraints. 
When we modify this minimal setup with the addition of a monomial GNMDC function with an appropriate amplitude, it inevitably shifts the values of both $n_s$ and $r$ from their optimal best-fit values. We study the Starobinsky setup in presence of the GNMDC monomial function of the form 
$\theta_0(\phi)=A_0 \phi^4$, and as we can see from figure \ref{ps}, increasing the strength of the monomial term leads to rendering the model inconsistent with the CMB observations. From our earlier discussion in section \ref{our-approach}, the presence of the monomial term in the GNMDC function is essential for a consistent model building and thus, we find that the Starobinsky model together with GNMDC is not a suitable scenario to work with.

\subsubsection{The hilltop-quartic model}
\label{hilltop-choice}

The hilltop inflationary scenarios are small field models in which inflation takes place near the maxima of the potential. These models are described by the potentials of the form
\beq
V(\phi)=V_0 \left[1-\left(\frac{\phi}{\mu}\right)^p\right].
\eeq
 According to Planck 2018 observations \cite{Planck:2018jri}, the quartic potential ($p=4$) provides a better fit to the data as compared to the quadratic potential ($p=2$). Compared to the single-parameter models such as the quadratic or the Starobinsky model, this  class of models is a two-parameter family of models which contain an extra parameter $\mu$. In our setup, we start with the quartic potential with $p=4$ and exploit the freedom of varying $\mu$. This allows us comparatively large ranges for $n_s$ and $r$, given that these models are well defined only in the limit $\phi < \mu$ \cite{Martin:2013tda}.

As we can see in figure \ref{ps}, in the minimal case, increasing $\mu$ simultaneously increases both $n_s$ and $r$, and the scenario becomes consistent with the  Planck + BICEP data only for $\mu \gg 1 $. We further study the behavior of this model in the presence of a monomial GNMDC term $\theta_0(\phi)$. 
Since the hilltop quartic model belongs to the class of small field models, we work with a negative index of the monomial term, i.e. we choose $\theta_0(\phi) =A_0 \phi^{-4}$, following the arguments of section \ref{g5phi0}. The presence of this term shifts $n_s$ to a higher value than its value in the minimal setup for a given $\mu$ by tuning $A_0$, as shown in figure \ref{ps}. Thus, to get optimal values of $n_s$ and $r$ in the presence of $\theta_0(\phi)$, we need to start with $\mu$ which corresponds to a smaller value of $n_s$ in the minimal setup. Henceforth, in our modified setup, we employ an extra advantage of the monomial GNMDC function, i.e. to fix the $n_s$ issue, thereby making the hilltop-quartic scenario viable, even in the regime $\mu < 1$. It is evident from figure \ref{ps} that for small $\mu$ if we increase the amplitude of the monomial term, the value of $n_s$ increases, but $r$ does not show a significant increment, and we arrive at the optimal values for $n_s$ and $r$ which are consistent with the observational bounds of Planck + BICEP datasets.

Depending on the value of $\mu$, the amplitude of the monomial term can be determined. The values of $\mu$ and $\theta_0(\phi)$  can be chosen appropriately to obtain the desired values of both $n_s$ and $r$. It is important to note that, the hilltop-quartic model leads to a degeneracy between the potential parameter $\mu$ and the GNMDC parameter $A_0$, as is also evident from figure \ref{ps}. It is possible to fix the value for any one of them while leaving the value of the other parameter to be determined from the best-fit parameter estimation methods. Moreover, this value will also depend on which datasets are being considered for the best-fit parameter estimation. For these values, the parameters of $\theta_1(\phi)$ can be adjusted such that $\theta(\phi) \equiv \theta_0(\phi) + \theta_1(\phi) > 0$. Finally, with these conditions in the GNMDC setup and staying consistent with the CMB data, one can obtain appropriate suppression and localized wiggles in the scalar power spectra. 


\section{Data analysis setup and methodology}
\label{data-analysis-and-methodology}

In this section, we shall discuss various details of our approach toward the numerical calculation of the primordial power spectrum, the methodology of our data analysis, and the details of the datasets that we use for obtaining the best-fit parameters constraints and the priors on various model parameters. 

\begin{figure}[!]
\centering
\includegraphics[width=0.69\columnwidth,height=16.6cm]{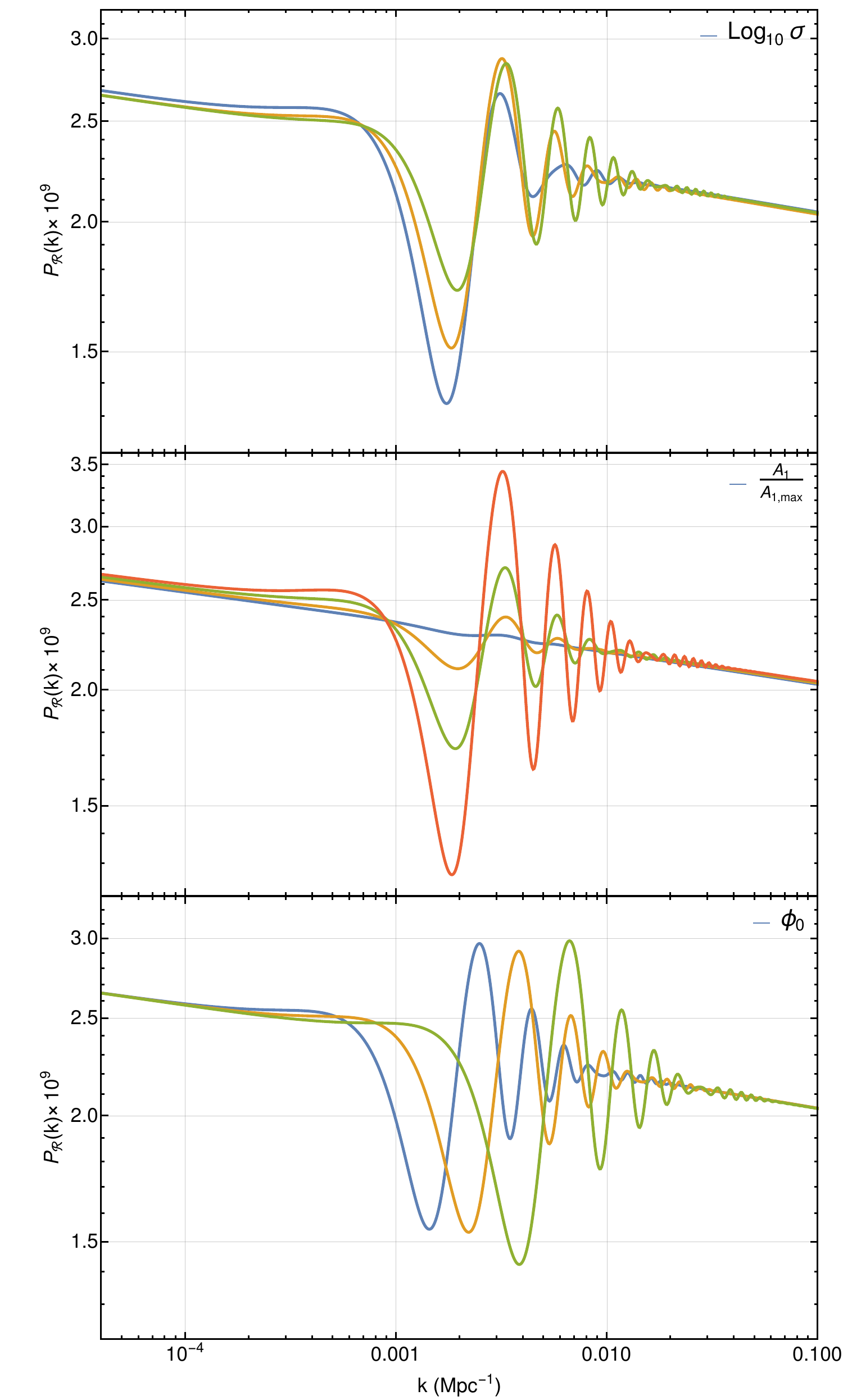}
\vskip 10pt
\caption{Primordial power spectrum of scalar perturbations for our model.
For each of the panels, we have varied one parameter while keeping the other two fixed. These three parameters viz. ($\sigma$, $A_1$, $\phi_0$) of the local term $\theta_1(\phi)$, distinctively affect the width, amplitude and the location of the features in the primordial power spectra, respectively. Note that, the GNMDC term is only responsible for generating such localized features in the spectrum. Away from the features, the spectrum is otherwise nearly scale-invariant or featureless.  These localized superimposed oscillations do lead to a better fit to the data as compared to a nearly scale-invariant power spectrum.}
\label{ps-diff}
\end{figure}

\subsection{Primordial power spectrum}
\label{pps}
The GNMDC scenario discussed in the previous section brings about  interesting and relevant oscillatory features in the primordial scalar power spectrum. In order to capture these  oscillations precisely, we have developed a robust numerical code that solves the complex GNMDC equations for the evolution of the background and perturbation equations to yield the primordial scalar power spectrum. As discussed earlier, we employ the hilltop-quartic model as our base inflationary model. While the monomial term $\theta_0(\phi)$ broadly controls the value of $n_s$ and $r$ at the pivot scale, the  localized GNMDC term i.e. $\theta_1(\phi)$ is largely responsible for generating features i.e. strong dip/suppression together with superimposed localized oscillations on very large scales. 
Our scenario has  inflationary potential parameters $\mu$, $V_0$, and GNMDC parameters $A_0$, $A_1$, $\phi_0$ and $\sigma$. Further, the maximum amplitude of the local GNMDC term $A_1$ is constrained to $A_{1,{\rm max}}$ from the requirement of an overall positive $\theta(\phi)$ as discussed in section \ref{our-approach}. 

In our numerical module, we integrate all the background and perturbation equations exactly without incorporating any approximations. 
For the background evolution, we solve the Friedmann and Klein-Gordon equations using the slow roll initial conditions, following our discussion of section  \ref{horndeski} and appendix \ref{appendix-1}. This requires specifying only the initial value of the inflaton field i.e. $\phi_i=\phi(N=N_i)$ which is constrained by requiring that inflation lasts for a long enough duration. 
Further, to obtain the primordial power spectrum, we evolve the scalar perturbations from the sub-horizon regime till they become frozen on the super-horizon scales  and impose the standard Bunch-Davies vacuum initial conditions in the sub-Hubble regime $k \gg aH$.
In all our numerical analyses, we ensure that we avoid the gradient instability and superluminality associated with the scalar perturbations. For the choice of the inflaton potential and the GNMDC coupling function in our setup, we find that the tensor perturbations remain exceedingly small ($r \sim 10^{-9}$) and thus, we do not take them into account in the data analysis. 
 The three parameters viz. ($\phi_0$, $\sigma$, $A_1$) of the local term $\theta_1(\phi)$, distinctively affect the location, width and the amplitude of the features in the primordial power spectra, respectively. A glimpse of the primordial power spectra of scalar perturbations obtained for different parameters of our model is shown in figure \ref{ps-diff} and for each panel, we have varied one parameter while keeping the  other two fixed.

As discussed earlier in section \ref{hilltop-choice}, there exists a degeneracy between the potential parameter $\mu$ and $\theta_0(\phi)$-parameter $A_0$. Thus, we fix the value of $\mu=0.09\, \Mpl$ and vary $A_0$ to achieve optimal fitting from MCMC chains. This leaves us with five model parameters to sample: $A_0$, $\phi_0$, $A_1$, $\sigma$, and $V_0$. The parameters $V_0$ and $A_0$ respectively fix the amplitude and the slope of the power spectrum around the pivot scale, $\phi_0$ determines the location of the suppression and oscillations in the power spectra, $\sigma$ localizes the features to the desired scales and $A_1$ controls the amplitude of the local GNMDC term.

\subsection{Datasets used in our analysis}

\begin{table}[t!]
\centering
\renewcommand{\arraystretch}{1.25}
\begin{tabular}{ |c|c| } 
\hline
Datasets & Likelihoods used \\
\hline
\multirow{1}{*}{Dataset I} & lowT+lowE+TT-Plik \\ 
\multirow{1}{*}{Dataset II} &
lowT+lowE+TTTEEE-Plik \\ 
\multirow{1}{*}{Dataset III} &
lowT+lowE+TT-Camspec12.5 \\ 
\multirow{1}{*}{Dataset IV} &
lowT+lowE+TTTEEE-Camspec12.5 \\ 
\hline
\end{tabular}
\vskip 10pt
    \caption{The four different likelihoods used in our analysis are characterized by four dataset combinations.}
    \label{tab:Table1}
\end{table}
Heretofore, we have discussed the generation of large-scale features in the scalar power spectrum due to the GNMDC term and outlined various issues and their resolutions for consistent model building. The next step is to understand how such a model compare with the observational datasets and obtain the best-fit parameter constraints on the GNMDC parameters along with the background cosmological parameters.
We use the latest release of the publicly available Planck 2018 CMB temperature and polarization (E-mode) anisotropy datasets \cite{Planck:2019nip}. Depending on the angular scales of interest and the distinguishable oscillatory features of a given model, one can choose the likelihoods provided by Planck for both small and large scales. Since our model primarily focuses on the CMB anomalies on the very large scales (low multipoles), we use the temperature and polarization likelihoods for these scales covering multipoles $\ell$ = $2 - 29$ in all the cases of our analysis. These likelihoods are \texttt{commander\_dx12\_v3\_2\_29} and \texttt{simall\_100x143\_offlike5\_EE\_Aplanck\_B} for the temperature and polarization anisotropies, respectively. From now on, we will refer to them as \texttt{lowT} and \texttt{lowE}, where low indicates the low multipoles ($\ell$ = $2-29$), and T/E for temperature/polarization data respectively. To probe the features of our GNMDC model at higher multipoles (i.e. on smaller scales), we use the Planck official likelihoods named \texttt{plik_rd12_HM_v22_TT} and \texttt{plik_rd12_HM_v22b_TTTEEE}  and we refer to them as \texttt{TT-Plik} and \texttt{TTTEEE-Plik} respectively.  Furthermore, to test the robustness of our results, we also use the recent \texttt{Camspec-12.5HMcl} likelihoods for high-$\ell$, which are obtained after reanalysis of the Planck 2018 data and span a slightly different range of multipoles. We refer to these as \texttt{TT-Camspec12.5} and \texttt{TTTEEE-Camspec12.5} respectively. We work with four dataset combinations which are listed in table \ref{tab:Table1}. These are \texttt{lowT+lowE+TT-Plik/Campec12.5} and \texttt{lowT+lowE+TTTEEE-Plik/Campsec12.5}, taking \texttt{lowT} and \texttt{lowE} common in all the cases.

\subsection{Model parameters sampling}

For a sampling of model parameters, we employ the publicly available code  \texttt{CosmoMC} which performs a detailed Markov Chain Monte Carlo (MCMC) analysis so as to find the best-fit values of various model parameters. The choice of working with the standard Metropolis Hasting sampling technique rather than nested sampling is due to the complicated numerical module to calculate the power spectrum in a GNMDC setup, which otherwise runs into difficulty when working with the latter.
The code for calculating the CMB power spectrum,       
\texttt{CAMB} \cite{Lewis:1999bs}, is contained within  \texttt{CosmoMC} which takes as input the background cosmological parameters including the parameters of the inflationary model, to yield the CMB temperature and polarization power spectra. 

\begin{table}[!]
\centering
\renewcommand{\arraystretch}{1.25}
\begin{tabular}{ |c|c|c|c| } 
\hline
Base and model parameters & Parameter & Priors \\
\hline
\multirow{4}{*}{$\Lambda$CDM parameters} &  $\Omega_bh^2$ & [0.005, 0.1]  \\ 
& $\Omega_ch^2$ & [0.001, 0.99] \\ 
& $\tau$ & [0.01, 0.8] \\ 
& $100\Theta_s$ & [0.5, 10] \\
\hline
\multirow{5}{*}{GNMDC parameters} &
$A_0$ & [21.4, 22.0] \\ 
& $\phi_0/ \phi_i$ & [1.05, 1.055] \\ 
& $\log_{10}\sigma$ & [-26.0, -24.0]   \\ 
& $V_0\times10^{17}$ & [5.9, 6.17] \\
& $A_1/A_{\rm 1,max}$ & [0.01, 0.7] \\
\hline
\end{tabular}
\vskip 10pt
\caption{The priors for the cosmological and model parameters used in our analysis. Note that, the range of priors remains the same for all four datasets.}
    \label{tab:Table2}
\end{table}

In the standard $\Lambda$CDM scenario which is our base reference model,  there are six cosmological parameters $\Omega_b, \Omega_c, \tau, \Theta_s, A_s, n_s$ where the first four are baryon density, cold dark matter density, optical depth at reionization and sound horizon respectively, while the last two are the inflationary parameters; $A_s$ setting the power spectrum normalization at the pivot scale and $n_s$ determining the slope of the primordial scalar power spectrum.
As we discussed earlier, in our set-up, instead of $A_s$ and $n_s$, we have  $V_0$, $A_0$, $A_1$, $\phi_0$ and $\sigma$, and hence, a total of nine parameters, including four cosmological parameters. In order to generate the theoretical CMB power spectra for various model parameters, we couple our primordial power spectrum module to \texttt{CAMB}, such that it takes as input the GNMDC inflation model parameters along with the four background cosmological parameters. 

In order to understand the improvement from our GNMDC inflationary scenario, we compare it with the reference $\Lambda$CDM models using \texttt{CosmoMC} with different datasets. 
The comparison is obtained in terms of the $\Delta\chi^2$ defined as $\Delta\chi^2 =  \chi^2_{\rm {\Lambda CDM}} - \chi^2_{\rm GNMDC}$.  
To further find the true set of model parameters that best fit the observational data and the corresponding $\Delta\chi^2$, we use the likelihood maximizer algorithm \texttt{BOBYQA} \cite{Powel:2009bob} available in \texttt{CosmoMC} setup which further spans the parameter space, obtained after the MCMC sampling and identifies the better likelihood regions in our samples. Such a  detailed analysis allows us to visualize the improvement in the fit coming from different scales in the power spectra. Finally, to correctly understand the improvement in the fit to each data set, we write down the total $\Delta\chi^2$ as follows
\begin{equation}
\Delta\chi^2_{\rm total} = \Delta\chi^2_{\rm low T} + \Delta\chi^2_{\rm low E} + \Delta\chi^2_{\rm high-\ell}+ \Delta\chi^2_{\rm prior}.
\end{equation}
\begin{figure}[!]
\centering
\includegraphics[width=0.98\columnwidth,height=20cm]{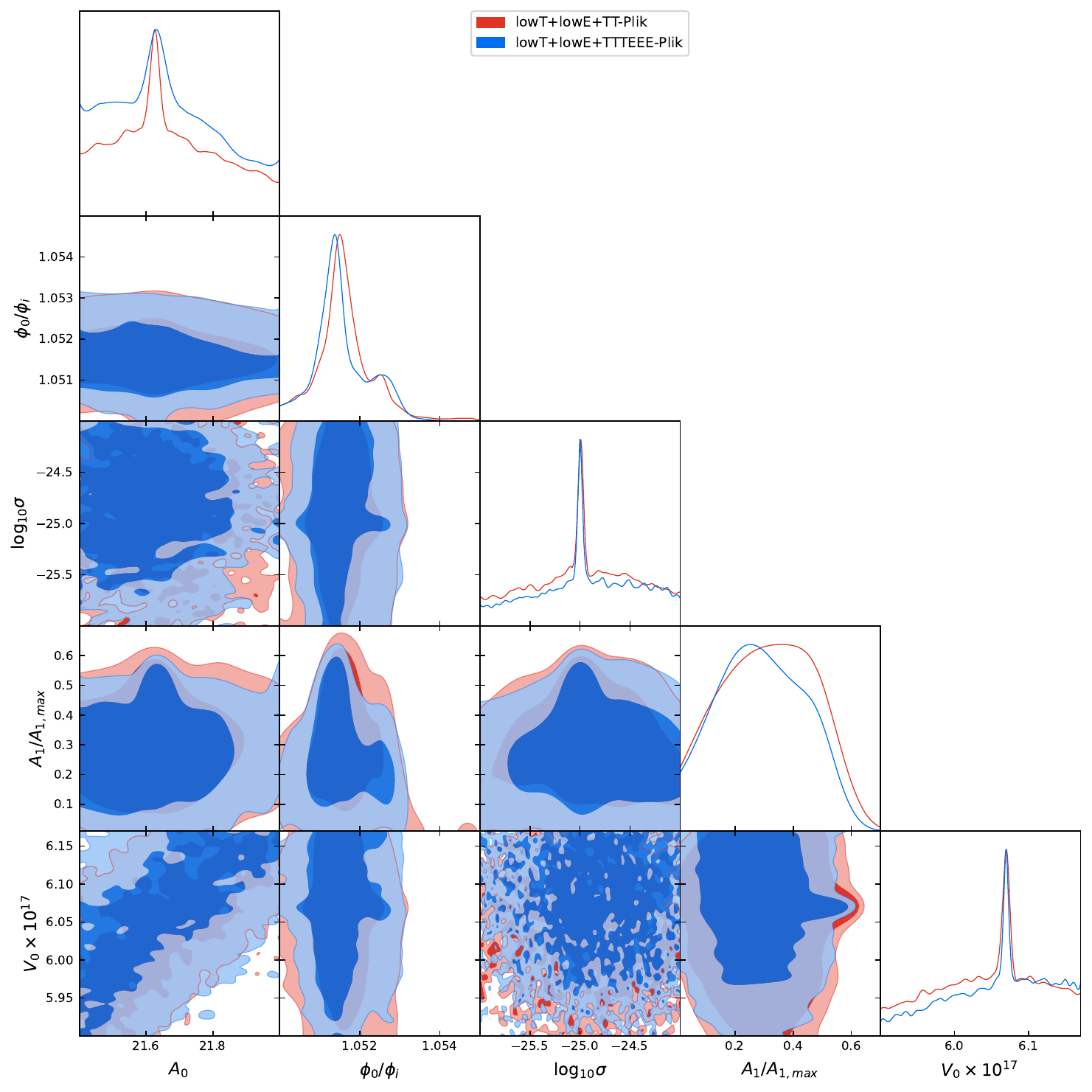}
\vskip 10pt
\caption{The triangle plot of the 1D and 2D posterior distribution obtained from the \texttt{CosmoMC} sampling using the \texttt{lowT+lowE+TT-Plik} and \texttt{lowT+lowE+TTTEEE-Plik} likelihoods for the five model parameters of our GNMDC set-up (excluding the four cosmological parameters). The plot indicates that the data can place tight constraints on various model parameters.}
\label{result-plik}
\end{figure}
\begin{figure}[!]
\centering
\includegraphics[width=0.98\columnwidth,height=20cm]{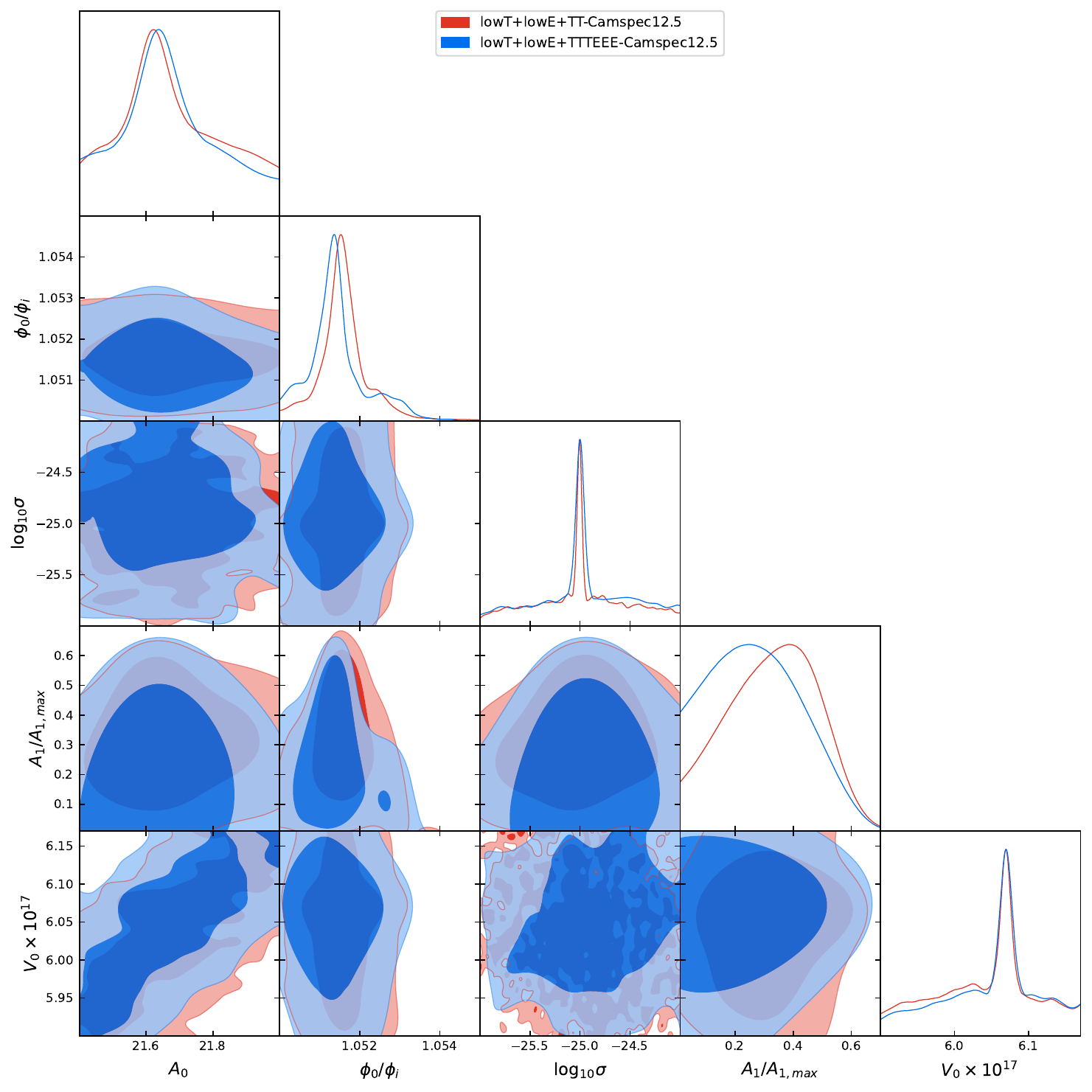}
\vskip 10pt
\caption{The triangle plot of the 1D and 2D posterior distribution obtained from the \texttt{CosmoMC} sampling using the \texttt{lowT+lowE+TT-Camspec12.5} and \texttt{lowT+lowE+TTTEEE-Camspec12.5} likelihoods for the five model parameters for our GNMDC set-up (excluding the four cosmological parameters). The plot indicates that the data can provide stringent constraints on various model parameters.}
\label{result-camspec}
\end{figure}
In our analysis, we explicitly provide the variation of each of the terms with respect to the base $\Lambda$CDM model for all the four datasets that we have considered. Note that, the first three terms correspond to the improvement arising from the comparison with different datasets while $\Delta\chi^2_{\rm prior}$ arises due to the priors on various model and nuisance parameters.
These nuisance parameters are additional parameters contained in the high-$l$ likelihoods for calibration of the data.

\subsection{Priors for the model parameters}
The effective prior volume in the multi-dimensional parameter space which is sampled from the distribution during the \texttt{CosmoMC} analysis, is specified by the range of priors on various cosmological and model parameters. Moreover, the optimal range of priors also plays a crucial role in determining the best-fit values of various model parameters. For the case of our GNMDC scenario, the priors on various cosmological and model parameters are summarised in Table \ref{tab:Table2}. Among these,  there are four base cosmological parameters, corresponding to the  $\Lambda$CDM model and five model parameters, describing the GNMDC scenario. 
In sec. \ref{pps}, we have already discussed how these model parameters affect the location and amplitude of various features arising in the power spectrum. We choose these parameter priors carefully and appropriately such that we capture the essential features in the power spectrum and also do not inflate the prior volume of the parameter space else the convergence take a large amount of time. 
Further, in our analysis, the range of priors remains the same for all four data combinations as outlined in Table \ref{tab:Table2}.

\section{Results}
\label{results}

\begin{table}[!]
\begin{center}
\renewcommand{\arraystretch}{1.5}
\begin{tabular}{ |p{2cm}||p{2.9cm}|p{2.9cm}|p{2.9cm}|p{2.9cm}| }
 \hline

 \hline
 Parameters & Dataset I &Dataset II & Dataset III & Dataset IV \\
 \hline

$\Omega_bh^2$   & $0.02219\pm 0.00017$  & $0.02238\pm 0.00013 $  &$0.02220\pm 0.00019$   &  $0.02225\pm 0.00013  $ \\

$\Omega_ch^2$ & $0.1200\pm 0.0010  $  & $0.12022\pm 0.00084 $  & $0.1194\pm 0.0013 $ & $0.11998\pm 0.00099 $ \\

$\tau$ & $0.0535\pm 0.0061$ & $0.0541\pm 0.0059 $    &$0.0537^{+0.0062}_{-0.0051}$& $0.0536^{+0.0059}_{-0.0051}$\\
 
$100\Theta_s$    & $1.04087\pm 0.00041$& $1.04091\pm 0.00030$ &$1.04078\pm 0.00040 $& $1.04102\pm 0.00027 $\\
\hline
$A_0$ & $21.666^{+0.089}_{-0.24} $& $21.659^{+0.078}_{-0.25} $   &$21.68^{+0.14}_{-0.18}$&$21.67^{+0.13}_{-0.16} $ \\
  
$\phi_0/ \phi_i$ & $1.05164^{+0.00047}_{-0.00072}$ &  $1.0516^{+0.0011}_{-0.00074}$ &$1.05161^{+0.00040}_{-0.00057}$&$1.05146^{+0.00045}_{-0.00078}$\\
 
$\log_{10}\sigma$ &  $-24.96\pm 0.50$& $-24.93^{+0.81}_{-0.39} $     &$-24.99\pm 0.46 $& $-24.96\pm 0.45 $ \\
 
$V_0\times10^{17}$ & $6.047^{+0.092}_{-0.059}$& $> 6.04 $ &$6.043^{+0.063}_{-0.076}$& $6.050^{+0.060}_{-0.072}$ \\
 
$A_1/A_{1,{\rm max}}$ &$0.32^{+0.18}_{-0.16} $ &  $0.30\pm0.15 $    &$0.33^{+0.17}_{-0.14}$& $0.28^{+0.13}_{-0.20} $\\
 \hline
\end{tabular}
\vskip 10pt
    \caption{We present the best-fit constraints on various model parameters from the MCMC analysis of the GNMDC scenario for four dataset combinations. The mean values of all the parameters along with their 1-$\sigma$ constraints are listed here. We find that the best-fit values of model parameters are very close to each other for all the different datasets and likelihood combinations.}
    \label{tab:Table3}
\end{center}
\end{table}

%
%
%
%
%
%
%

In this section, we shall present and discuss in detail the results of our analysis. These include the posterior distribution of the parameters obtained from the MCMC sampling using \texttt{GetDist} \cite{Lewis:2019xzd}, the best-fit values of model parameters calculated using the \texttt{BOBYQA} routine, the detailed $\Delta\chi^2$ compared to the vanilla $\Lambda$CDM model with power-law power spectrum and finally, the best-fit residual CMB angular power spectra, showing significant improvement in fitting the Planck 2018 data.
For all these cases, the MCMC runs are analyzed after the convergence is reached for all the chains based on the Gelman-Rubin convergence criteria \cite{Gelman:1992zz} and we set $R-1 \sim 0.05$ for all our chains analysis.

The results of our data analysis are shown in figure \ref{result-plik} wherein we display the 1D and 2D posterior distribution of the sampled model parameters for the dataset combinations \texttt{lowT+lowE+TT-Plik} and \texttt{lowT+lowE+TTTEEE-Plik}. As is evident from the figure, the posteriors for both datasets nearly overlap with each other for all the parameters.
The posteriors also indicate that the data does not put  strong constraints on all five parameters. However, the theoretical priors combined with the observational datasets seem to indicate a preferred value for some parameters. As evident from figure \ref{result-plik} and \ref{result-camspec}, only the parameters $\phi_0$ and $A_1/A_{1,{\rm max}}$ seem to be well constrained, with nearly closed contours.

The parameter $\phi_0$, which determines the position of the prominent dip in the power spectrum, follows a near Gaussian profile, preferring a mean value corresponding to the dip-like feature around multipoles $\ell \sim 20-30$, thereby improving the fit in the CMB temperature power spectrum on large scales. Next, the parameter $A_1/A_{1,{\rm max}}$, which controls the amplitude of the localized GNMDC function, also admits a non-zero mean value indicating that the data indeed favors the presence of the local GNMDC term in our setup, which brings non-trivial features in the CMB power spectrum on relevant scales. The parameter $\sigma$, which controls the width of the feature, though shows a sharp peak corresponding to a preferred mean value but forms a rather scattered posterior. 
Further, the posteriors for the parameters $A_0$ and $V_0$, which collectively fix the values of $n_s$ and $r$ around the pivot scale, seem to be mostly prior bound. Indeed, there exists a degeneracy between $V_0$ and $A_0$, thereby, one can obtain better constraints on either of these, by fixing the other.
For the case of \texttt{lowT+lowE+TT-Camspec12.5} and \texttt{lowT+lowE+TTTEEE-Camspec12.5}, we observe very similar behavior in the posterior distribution, as shown in figure \ref{result-camspec}. There also exists consistency in the best-fit constraints among all the datasets that we have considered in our analysis.

\begin{table}[!]
\begin{center}
\renewcommand{\arraystretch}{1.4}
\begin{tabular}{ |p{2.5cm}||p{2.5cm}|p{2.5cm}|p{2.5cm}|p{2.5cm}| }
 \hline

 \hline
 Parameters & Dataset I &Dataset II &Dataset III &Dataset IV\\
 \hline
 
 $A_0$ &  21.5599& 21.5858& 21.6669 & 21.6008  \\
  
 $\phi_0/ \phi_i$  &  1.0515 &1.0513 & 1.0515& 1.0513\\
 
 $\log_{10} \sigma$ &  -25.5429 & -24.9896& -25.1201 & -24.7700 \\
 
 $V_0\times10^{17}$ &  5.9805 & 6.0604 & 6.0265 & 6.0072 \\
 
 $A_1/A_{1,{\rm max}}$ &  0.4770 & 0.4052 & 0.4743 & 0.3547\\
 \hline
 $\Delta\chi_{\rm lowT}^2$ & 5.3080  & 4.6490& 5.0740 & 4.6220 \\  
 $\Delta\chi_{\rm lowE}^2$ & -0.0110  & -0.5880 & -0.1160 & -0.1250 \\
 $\Delta\chi_{\rm high-\ell}^2$ & 1.27 & 1.90 & 2.3310 & 0.8140 \\
 $\Delta\chi_{\rm prior}^2$ & -0.1133 & 0.1596 & -0.4145 & 0.0036\\
 \hline\hline
$\Delta\chi^2_{\rm total}$ & 6.4537 & 6.1206 & 6.8745 & 5.3146\\
 \hline
\end{tabular}
\vskip 12pt
    \caption{The best-fit constraints on various model parameters obtained after the \texttt{BOBYQA} analysis for the different likelihood combinations. The bottom rows also list each contribution in   $\Delta\chi^2$ compared to the reference $\Lambda$CDM model and finally, the net improvement $\Delta\chi^2_{\rm total}$ for our scenario corresponding to different datasets.}
    \label{tab:Table4}
\end{center}
\end{table}

\begin{figure}[!]
\centering
\includegraphics[width=0.86\columnwidth,height=9cm]{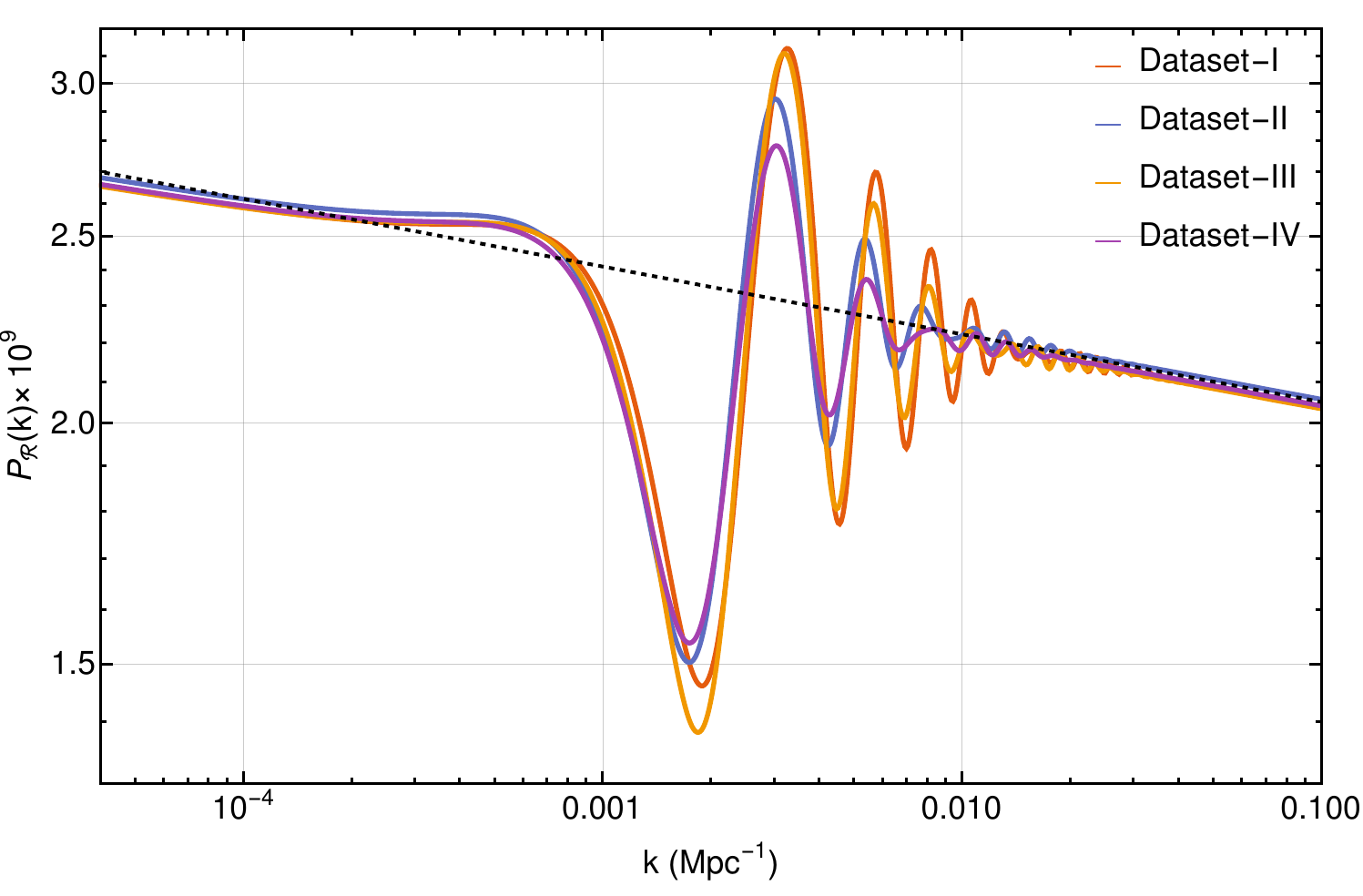}
\vskip 2pt
\caption{The scalar power spectra are plotted for the best-fit values of our GNMDC model parameters obtained after the \texttt{BOBYQA} analysis, along with a nearly scale-invariant spectrum for the best-fit values of $A_s$ and $n_s$ (in dashed). While the global features in the spectra remain the same, the different datasets seem to prefer these localized modulations with slightly different magnitudes and locations.}
\label{best-fit-ps}
\end{figure}

In table \ref{tab:Table3},  we have listed the mean values of all the model parameters along with the background parameters together with 1-$\sigma$ upper and lower bounds, from our data analysis for all four datasets. Although we have not listed the $\Lambda$CDM parameters explicitly, we do not find any significant shift in the background cosmological parameters in our analysis with respect to the $\Lambda$CDM model.

\subsection{Best-fit values of model parameters and the CMB power spectrum}
In order to obtain further improvement in the fit and the true best-fit values of model parameters, we use the 
\texttt{BOBYQA} routine
for all four datasets. 
In table \ref{tab:Table4}, we have listed the best-fit results obtained from \texttt{BOBYQA} for all the five parameters. 
We found that the best-fit values differ slightly from each other for the parameters of our model.
We further quote the obtained improvement in the fit, $\Delta\chi^2$, for individual likelihoods, as well as the total improvement, in comparison to the base $\Lambda$CDM model with the power-law power spectrum. 
\begin{figure}[!]
\vskip -40pt
\centering
\includegraphics[width=1.0\columnwidth,height=10.75cm]{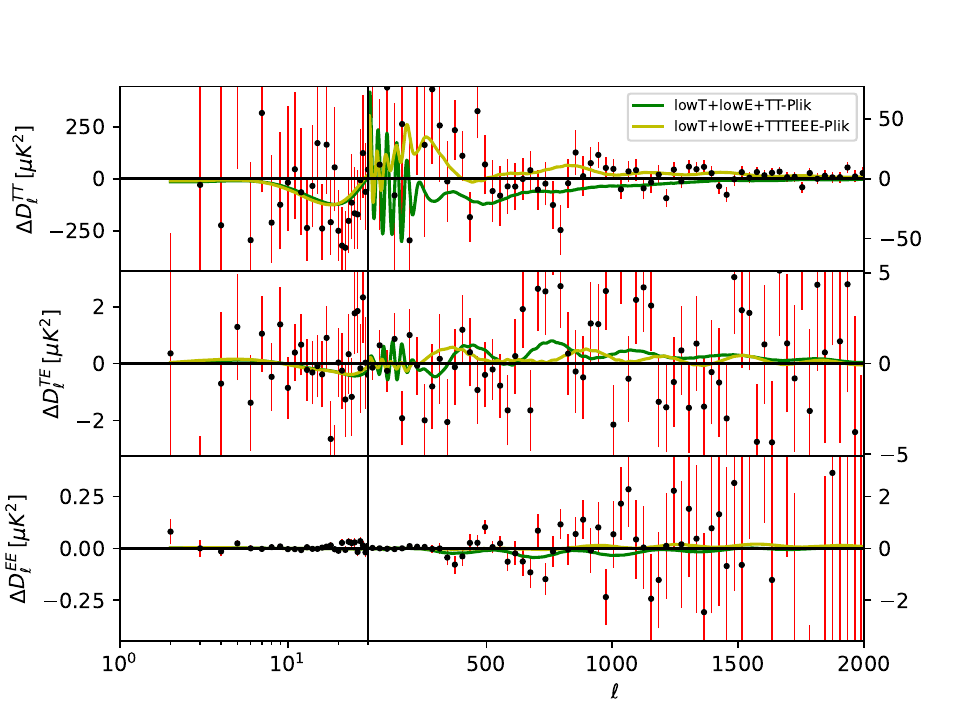}
\vskip -10pt
\includegraphics[width=1.0\columnwidth,height=10.75cm]{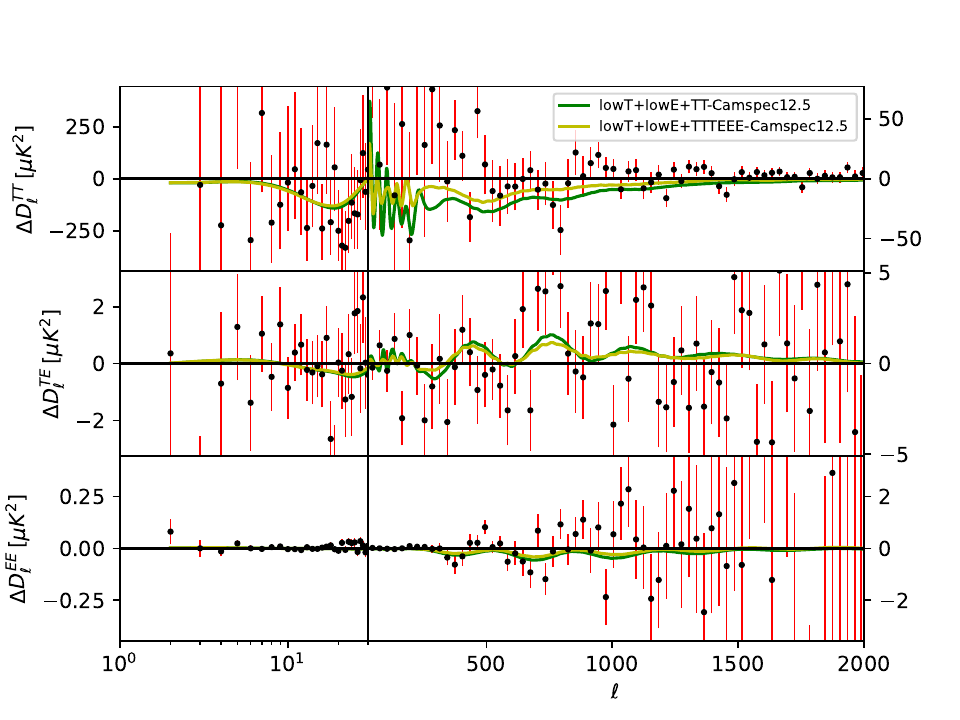}
\caption{The residual CMB angular power spectra are plotted for our GNMDC scenario with respect to the base $\Lambda$CDM model, corresponding to the best-fit parameters obtained from \texttt{BOBYQA} analysis for all the four  datasets (mentioned in table \ref{tab:Table4}). The improvement arising from the low-$\ell$ regime turns out to be more significant than the high-$\ell$ regime. 
}
\label{cmb-spectra}
\end{figure}

As evident from table \ref{tab:Table4} for all four datasets, we get a  significant improvement with $\Delta\chi^2 \sim 4.6-5.3$ from the lowT part i.e. from the CMB temperature anisotropy data on low multipoles. This is what we targeted while choosing the toy model for $\theta_1(\phi)$, as the improvement mostly arises from the prominent dip in the primordial power spectra corresponding to large scales caused due to $\theta_1(\phi)$ term.
There is no significant improvement coming from the lowE polarisation anisotropy data, indicating that the data is merely sensitive to the modifications in the power spectra introduced in our setup. 
In our best-fit samples, a small improvement to the fit on smaller scales (large multipoles) is also seen, $\Delta\chi^2 \sim 0.8-2.3$ for all four datasets. These features on smaller scales in the CMB data can be attributed to the decaying oscillatory features in the primordial power spectra following the most prominent dip, as also seen in figure \ref{ps-diff}. Although we mentioned that the improvement on smaller scales is not as significant, since the features in the power spectra are obtained to primarily focus on anomalies on large scales. Indeed, by some other suitable choice of coupling function $\theta_1(\phi)$, which induces small oscillations on relatively smaller scales in the power spectrum similar to \cite{Braglia:2021ckn, Braglia:2021rej}, one can obtain significant improvement from these scales too. Further the $\chi^2_{\rm prior}$ which arises from the nuisance parameters for the likelihoods is nearly the same as that of the reference model. Finally, in terms of the overall improvement, we find that $\Delta\chi^2_{\rm total} \sim 5.3 - 6.8$ and the maximum improvement in $\Delta\chi^2_{\rm total} \sim 6.8$ is obtained for the  dataset \texttt{lowT+lowE+TT-Camspec12.5}. On a final note, in all four dataset combinations, the dominant  improvement in $\Delta\chi^2_{\rm total}$ comes from the low multipole temperature anisotropies data, indicating that the modifications introduced in the primordial power spectrum in our setup, are well accommodated by the anomalous features in CMB on low multipoles ($\ell< 30$). This summarizes the best-fit improvement obtained for the GNMDC model that we have considered in our analysis. 

It might be very interesting to explore other coupling functions within the GNMDC set-up which might lead to an even better improvement in the lowT regime e.g. due to a suppressed power spectrum on very low multipoles so as to address the low quadrupole anomaly in the CMB as well as lead to other characteristic modulations at higher multipoles to get even better fit from the high-$\ell$ data.

In figure \ref{best-fit-ps}, we plot the scalar power spectra for the best-fit values of model parameters obtained after the  \texttt{BOBYQA} analysis for all four datasets combinations. While the broad behavior of the features in the spectra (a broad dip followed by localized  superimposed oscillations) remains the same, the different datasets still suggest a slight difference in the location and amplitude of these features. Away from these localized features, the power spectra reduce to a nearly scale-invariant power spectrum, as evident from the figure \ref{best-fit-ps}. It is interesting to point out that these features strongly resemble the features obtained in a canonical single field inflationary model with the introduction of a step in the inflaton potential \cite{Covi:2006ci, Hazra:2010ve, Hazra:2017joc}. Needless to say, our GNMDC scenario offers a new and interesting possibility to generate such large-scale features in the spectrum to prove to be a better fitting model to the CMB observations. It will be interesting to examine if the oscillatory features arising in our model also provide a resolution to the lensing anomaly \cite{Domenech:2019cyh, Domenech:2020qay}.

Further, to understand the resulting imprints in the CMB, we have plotted the residual CMB angular power spectra $\Delta D_{\ell}^{TT}$, $\Delta D_{\ell}^{TE}$ and $\Delta D_{\ell}^{EE}$ corresponding to the best-fit values of the model parameters for all the four datasets which are displayed in figure \ref{cmb-spectra}.
The top panel in this figure corresponds to datasets I and II, while the lower is plotted for
the datasets III and IV.
It is evident from this figure that all these best-fit CMB spectra show similar features. Our GNMDC scenario is not able to address the low quadrupole. Still, it results in oscillatory features around $\ell \sim 20 -30$, which improves the overall fit in the low-$\ell$ regime as compared to the base featureless model. Further, we do not get any significant improvement in the E-mode polarization autocorrelation data ($D_{\ell}^{EE}$), suggesting that such a setup may not improve the fit to the same beyond the standard model.
Notably, the improvement arising from the high-$\ell$ regime seems to somewhat differ from each other for all four datasets. Our scenario does generate tiny small-scale features but not as relevant and prominent as the clock signals, discussed in \cite{Braglia:2021sun, Braglia:2021rej}.
We close this section with an interesting prospect that the  class of GNMDC models that we have considered in this work, hold the promise of being viable alternatives of primordial feature models, and they can be further expanded by considering different forms of the GNMDC functions as well as other mechanisms to generate prominent small scale features. 

\section{Conclusions and discussions}
\label{conclusions}

Precision observations of CMB anisotropies over the last few decades are one of the most outstanding achievements of modern cosmology. These anisotropies contain crucial information about the evolution of the universe at the earliest epochs, particularly during the inflationary phase. Although the present observational data strongly favor the standard $\Lambda$CDM model as the concordance model of the universe, various anomalies present in the CMB data may nevertheless hint at some new primordial physics beyond the standard model of cosmology. Interestingly, these large-scale anomalies have always been present in the CMB observations from COBE, and WMAP to the most recent and precise Planck datasets, which makes them more intriguing to look at and understand their origins better. While they could arise due to foreground residuals and/or systematic effects, their origin could also be primordial, which could point towards non-trivial dynamics beyond the simplest inflationary models, which typically lead to a nearly scale-invariant power spectrum on all scales.

In this paper, we have studied the cosmological implications of the GNMDC term that arises within the framework of Horndeski theories and explored whether large-scale features in the CMB can be generated in this scenario. Since the evolution of background and perturbations is quite involved in this model, we have developed an accurate numerical module to compute the primordial spectra of scalar and tensor perturbations. Further, we have compared our model with the CMB anisotropies data from Planck in temperature (TT), E-mode polarization (EE), and their cross-correlation (TE) and explored the parameter space to find out the best-fit values of various model parameters. We found that this class of models can indeed generate large-scale features in the CMB angular power spectra for a specific choice of the GNMDC function and provide an improvement over the reference $\Lambda$CDM model with a featureless power-law primordial spectrum. The CMB angular power spectrum corresponding to the best-fit parameters indicates that the dominant improvement primarily arises from the low-$\ell$ region with marginal improvement from the high-$\ell$ region. 
To our knowledge, this is the first time that such models have been employed to understand the origin of large-scale features in the CMB angular power spectrum. Our results demonstrate the possibility of explaining large-scale features in the CMB by going beyond the canonical scalar field models without any additional features in the inflaton potential but within the single field inflationary framework. 

The GNMDC scenario offers a richer phenomenology in terms of freedom to choose the form of the coupling function. Hence, it might be interesting to explore if these models can also explain the presence of other feature anomalies in the CMB, e.g. low quadrupole suppression and peculiar clock features at smaller scales. As discussed earlier, from a theoretical perspective, a model can be a consistent physical model only if it avoids the gradient and Ostrogradsky instabilities.
However, from a statistical perspective, a model can be termed a better alternative only if it improves the fit to the data compared to a reference model. Needless to say, a preferable scenario would be the one that provides a satisfactory  explanation to multiple anomalies in the CMB observations and also proves to be a significantly better fit to the data at the expense of fewer additional parameters.
\begin{figure}[!]
\centering
\includegraphics[width=0.7\columnwidth,height=11cm]{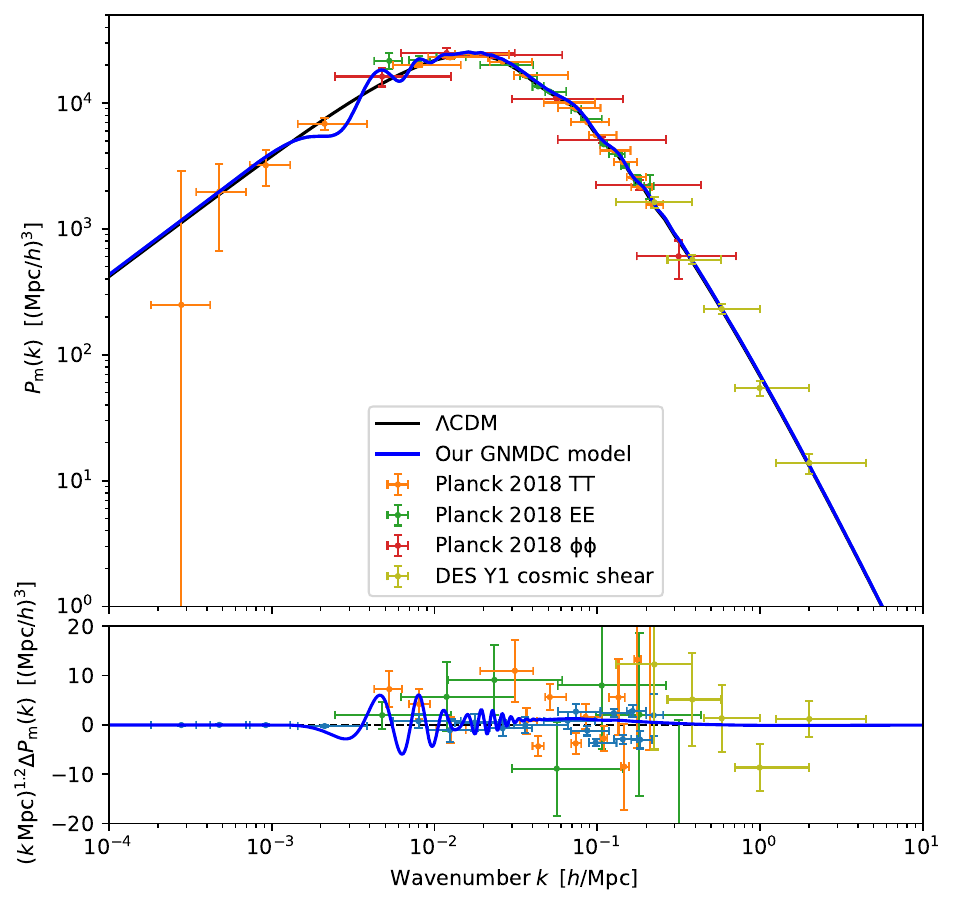}
\vskip 4pt
\caption{The linear matter power spectrum at $z=0$ (top) and the residuals (below) for our scenario for the best-fit values of model parameters.  Precise LSS observations contain the enormous potential to constrain features in the spectrum independently but not the features on very large scales as those scales remain  inaccessible by LSS \cite{Chabanier:2019eai}.}
\label{matter-ps}
\end{figure}
Within this GNMDC framework, it will also be helpful to work with different choices of the coupling function leading to suitable oscillations on smaller scales which can provide a better improvement arising from smaller scales.
In addition to the CMB anisotropies data from Planck, large-scale features in the power spectrum can, in principle, also be independently constrained by the 
large-scale structure (LSS) observations of galaxy surveys as the primordial features also leave distinct imprints in the matter power spectrum, as evident from figure \ref{matter-ps}. However, localized features on scales larger than $k \sim 0.01\, h\, {\rm Mpc^{-1}}$ may not be constrained by the LSS data while  features on smaller scales or running features on all scales can be independently constrained by the LSS observations \cite{Chen:2016vvw, Ballardini:2016hpi, Ballardini:2017qwq, Palma:2017wxu, LHuillier:2017lgm, Beutler:2019ojk, Ballardini:2019tuc, Esmaeilian:2020rbq, Li:2021jvz, Ballardini:2022wzu}. Upcoming LSS surveys such as Euclid and LSST should be able to provide better constraints on such primordial spectral features \cite{Debono:2020emh, Chandra:2022utq}.
Thus, LSS observations can be used as a complementary probe along with the CMB data to constrain non-trivial features in the power spectra and the deviations from scale invariance on sub-CMB scales. 
Moreover, future 21-cm tomographic observations also have enormous potential to constrain such primordial features to a great extent  \cite{Chen:2016zuu, Xu:2016kwz, Balaji:2022zur}.
In addition, spectral distortions of the CMB power spectrum provide an entirely complementary window on the scale dependence of the primordial power spectrum and small-scale features since they strongly depend on the spectral amplitude at scales smaller than LSS. 
Future experiments such as PIXIE \cite{Kogut:2011xw} or PRISM \cite{PRISM:2013fvg} should be able to provide interesting constraints on departures from a featureless
primordial power spectrum.
All these observations from CMB, LSS, 21-cm, and spectral distortions prove to be complementary. Together, they will provide stringent constraints on the primordial power spectrum over a wide range of scales.

Primordial non-Gaussianity (PNG) is also an important probe that can provide independent constraints on the scale dependence of the power spectrum. In particular, inflationary models with features in the power spectrum are well known to generate correlated signatures in PNG \cite{Chen:2006xjb, Chen:2008wn, Adshead:2011jq, Hazra:2012yn, Fergusson:2014hya, Fergusson:2014tza}. Therefore, present and future PNG constraints from CMB and LSS can constrain these feature models better. 
Near future CMB experiments aimed at precise measurements of CMB E- and B-mode polarization such as CMB-S4 \cite{CMB-S4:2016ple}, LiteBIRD \cite{LiteBIRD:2020khw,2019BAAS51g.286L}, Simons Observatory (SO) \cite{SimonsObservatory:2018koc, SimonsObservatory:2019qwx} and CMB-Bharat \cite{cmb-bharat} should also be able to provide relevant constraints on the small scale dynamics of inflationary models with features and their cross-correlations with other probes should further tighten the parameter constraints. With the availability of future precise data, many consistent models may be ruled out or strongly constrained. The prospects of a better understanding of the interconnection between primordial spectral features and underlying physics will undoubtedly be improved. 
Therefore, future cosmological observations offer an exciting promise to constrain the primordial universe better.

\acknowledgments
We would like to thank Dhiraj K. Hazra for very fruitful discussions on various aspects of cosmological data analysis and L. Sriramkumar for useful comments on the draft. We acknowledge the use of high-performance computational facilities at the Supercomputer Education and Research Centre (SERC) of the Indian Institute of Science, Bengaluru, India. RKJ wishes to acknowledge financial support from the new faculty seed start-up grant of the Indian Institute of Science, Bengaluru, India, Science and Engineering Research Board, Department of Science and Technology, Government of India, through the Core Research Grant~CRG/2018/002200 and the Infosys Foundation, Bengaluru, India through the Infosys Young Investigator award.

\appendix
\section{GNMDC equations for background and  perturbations}
\label{appendix-1}
In this appendix, we present the GNMDC equations for the background evolution and the scalar and tensor perturbations in terms of efolds $N$, which we employ in our numerical code. 

\subsection{Background equations}

The Friedmann equations for our GNMDC setup, as given in Eqn. (\ref{frw1}) and (\ref{frw2}), can be written in terms of $N$  as

\bea 
3 H^2&=&\kappa^2\l[\frac{1}{2} H^2 \phi_N^2 \left(9 H^2 \theta (\phi )+1\right)+V(\phi )\r],
\\
-2 H H_N&=&\kappa^2\bigg[\l(1+3\kappa^2\theta(\phi)H^2-\kappa^2\theta(\phi)H H_N\r) H^2 \phi_N^2-\kappa^2\theta'(\phi)H^4\phi_N^3 \bigg.  \nonumber \\
\bigg.
&-& 2\kappa^2\theta(\phi)H^2\phi_N\left(H^2 \phi_{NN}+H H_N \phi_N\right)\bigg],
\eea
and the Klein-Gordon equation becomes
\bea
\left[1+3 H^2 \theta (\phi )\right] \left(H^2 \phi_{NN}+H H_N \phi_N\right)+3 H^2 \phi_N \left[\left(3 H^2+2 H H_N \right) \theta
   (\phi )+1\right]  \nonumber \\
   +\frac{3}{2} H^4 \phi_N^2 \theta '(\phi )+V'(\phi)=0,
\eea
where $\phi_N={d\phi}/{dN}$, $\phi_{NN}={d^2\phi}/{dN^2}$, $H_N={dH}/{dN}$ and $\theta'(\phi)={d \theta(\phi)}/{d\phi}$. We set $\kappa=1$ in our numerical code. 

\subsection{Scalar and tensor perturbations}

The Mukhanov-Sasaki equation for the Fourier modes of curvature perturbations $\mathcal{R}_k$, as given in eqn. (\ref{rk-eqn}), can be further written in terms of $N$ as follows
\beq
\frac{d^2\mathcal{R}_k}{dN^2}+\left(1+\frac{H_N}{H}+\frac{2z_{s_N}}{z_s}\right)\frac{d\mathcal{R}_k}{dN}+\left(\frac{k^2c_s^2}{a^2H^2}\right)\mathcal{R}_k=0 \, .
\label{rk-N}
\eeq
Here, $z_{s_N}={dz_s}/{dN}$ and $c_s^2$ and $z_{s_N}/z_s$ are given by

\begin{align}
\begin{split}
c_s^2  &= \frac{1}{\left (\left(H^2 \theta (\phi) \phi_N^2 -2\right) \left(9 H^4 \theta (\phi)^2 \phi_N^2-H^2 \theta (\phi) \left(\phi_N^2-6\right)+2\right)^2 \right)}\\
&\times \Big[ H^4 \theta (\phi)^3 \phi_N^2 \left(-6 H^4 \phi_N^5 \theta '(\phi)+H^2 \left(3 \phi_N^4-244 \phi_N^2+420\right)-48 \phi_N
   V'(\phi)\right) \\
   & +2 \theta (\phi)^2 \left(12 H^6 \phi_N^5 \theta '(\phi)+24 H^2 \phi_N V'(\phi)+H^4 \left(-7 \phi_N^4+100 \phi_N^2
   -36\right)\right)\\
   & -4 H^2 \theta (\phi) \left(6 H^2 \phi_N^3 \theta '(\phi)-5 \phi_N^2+12\right)+6 H^6 \theta (\phi)^4 \phi_N^4
   \left(13 H^2 \left(\phi_N^2-9\right)+2 \phi_N V'(\phi)\right)\\
   & +351 H^{10} \theta (\phi)^5 \phi_N^6-8 \Big]
\end{split}
\label{cs-N}
\end{align}

\begin{align*}
\begin{split}
\frac{z_{s_N}}{z_s} &= 
\frac{1}{\left (H^2 \phi_N \left(H^2
   \phi_N^2 \theta (\phi )-2\right) \left(3 H^2 \phi_N^2 \theta (\phi )-2\right) \left(9 H^4 \phi_N^2 \theta (\phi )^2-H^2
   \left(\phi_N^2-6\right) \theta (\phi )+2\right)^2 \right)}\\
	&\times \Big[ H^2 \phi_N \left(3 H^8 \phi_N^4 \theta (\phi )^3 \left(\phi_N^5 \theta '(\phi )+\left(-20 \phi_N^4+54
   \phi_N^2+936\right) \theta (\phi )\right) \r. \Big. \\
   &  -H^6 \phi_N^2 \theta (\phi )^2 \left(2 \left(5 \phi_N^2+24\right) \phi_N^3
   \theta '(\phi )+\left(3 \phi_N^6-220 \phi_N^4-132 \phi_N^2+720\right) \theta (\phi )\right)\\
   &  +2 H^4 \theta (\phi ) \left(2 \left(\phi_N^2+24\right) \phi_N^3 \theta '(\phi )+\left(7 \phi_N^6-132 \phi_N^4
   +60 \phi_N^2  -144\right) \theta (\phi )\right)\\
   &  -4 H^2 \left(\left(5 \phi_N^4-36 \phi_N^2+48\right) \theta (\phi )-2 \phi_N^3
   \theta '(\phi )\right)+972 H^{12} \phi_N^8 \theta (\phi )^6 \\
   &  + \l.  27 H^{10} \phi_N^6 \left(\phi_N^2-84\right) \theta (\phi )^5+8
   \left(\phi_N^2-4\right)\right)\\
   & +V'(\phi ) \left(27 H^{10} \phi_N^8 \theta (\phi )^5-9 H^8 \phi_N^6 \left(\phi_N^2
   +12\right) \theta (\phi )^4+8 H^6 \phi_N^4 \left(5 \phi_N^2+24\right) \theta (\phi )^3 \r. \\
   &  \Big. - \l. 8 H^4 \phi_N^2 \left(7 \phi_N^2+18\right) \theta (\phi )^2+16 H^2 \left(2 \phi_N^2-3\right) \theta (\phi )-16\right) \Big]
\end{split}
\end{align*}
The Fourier mode equation for the tensor perturbations $h_k$ in terms of $N$ can be written as
\beq
\frac{d^2h_k}{dN^2}+\l(1+\frac{H_N}{H}+\frac{2 z_{T_N}}{z_T}\r)\frac{dh_k}{dN}+\left(\frac{c_T^2k^2}{a^2H^2}\right)h_k=0,
\label{hk-N}
\eeq
where
\beq
c_T^2=\frac{1+H^2 \theta (\phi) \phi_N^2/2}{1- H^2 \theta (\phi) \phi_N^2/2}
\eeq
and
\begin{align*}
\begin{split}
\frac{z_{T_N}}{z_T} &= 
\frac{1}{\Big(2 \left(H^2 \theta (\phi) \phi_N^2-2\right) \left(9 H^4 \theta (\phi)^2 
   \phi_N^2-H^2 \theta (\phi) \left(\phi_N^2-6\right)+2\right)\Big)} \\ 
   & \times \Big[2 H^2 \left(\phi_N^3 \theta '(\phi)+\theta (\phi)^2 \phi_N^3 V'(\phi)-2 \theta (\phi) \left(\phi_N^2+6\right)\right)\Big. \\
   & \Big. +H^4 \theta(\phi) \phi_N^2 \left(4 \theta (\phi) \left(4 \phi_N^2-15\right)-\phi_N^3 \theta '(\phi)\right)+72 H^6 \theta (\phi)^3 \phi_N^4-4
   \left(\theta (\phi) \phi_N V'(\phi)+2\right)\Big] \, .
\end{split}
\end{align*}

As mentioned earlier, the mode equation (\ref{rk-N}) for $\mathcal{R}_k$ is evolved for each Fourier  mode with the Bunch-Davies initial conditions imposed on $\mathcal{R}_k$ and its derivative when the modes are deep inside the horizon (subhorizon regime, i.e., $k/{aH} \gg 1$). In terms of conformal time $\tau$, we can write
\begin{equation*}
    \mathcal{R}_k(\tau)\Big\vert_{k \gg aH}=\frac{1}{z_s}\frac{e^{-i c_s k\tau}}{\sqrt{2c_s k}}\; .
\end{equation*} 
Similarly, the Bunch-Davies initial conditions for the tensors on $h_k$ are similar to those of scalars with $c_s \to c_T$ and $z_s \to z_T$.

\bibliographystyle{JHEP}
\bibliography{references.bib}

\providecommand{\href}[2]{#2}\begingroup\raggedright\begin{thebibliography}{100}

\bibitem{Starobinsky:1980te}
A.~A. Starobinsky, {\it {A New Type of Isotropic Cosmological Models Without
  Singularity}},  {\em Phys. Lett. B} {\bf 91} (1980) 99--102.

\bibitem{Guth:1980zm}
A.~H. Guth, {\it {The Inflationary Universe: A Possible Solution to the Horizon
  and Flatness Problems}},  {\em Phys. Rev. D} {\bf 23} (1981) 347--356.

\bibitem{Linde:1981mu}
A.~D. Linde, {\it {A New Inflationary Universe Scenario: A Possible Solution of
  the Horizon, Flatness, Homogeneity, Isotropy and Primordial Monopole
  Problems}},  {\em Phys. Lett. B} {\bf 108} (1982) 389--393.

\bibitem{Albrecht:1982wi}
A.~Albrecht and P.~J. Steinhardt, {\it {Cosmology for Grand Unified Theories
  with Radiatively Induced Symmetry Breaking}},  {\em Phys. Rev. Lett.} {\bf
  48} (1982) 1220--1223.

\bibitem{Linde:1983gd}
A.~D. Linde, {\it {Chaotic Inflation}},  {\em Phys. Lett. B} {\bf 129} (1983)
  177--181.

\bibitem{Starobinsky:1979ty}
A.~A. Starobinsky, {\it {Spectrum of relict gravitational radiation and the
  early state of the universe}},  {\em JETP Lett.} {\bf 30} (1979) 682--685.

\bibitem{Mukhanov:1981xt}
V.~F. Mukhanov and G.~V. Chibisov, {\it {Quantum Fluctuations and a Nonsingular
  Universe}},  {\em JETP Lett.} {\bf 33} (1981) 532--535.

\bibitem{Tegmark:2004qd}
M.~Tegmark, {\it {What does inflation really predict?}},  {\em JCAP} {\bf 04}
  (2005) 001, [\href{http://arxiv.org/abs/astro-ph/0410281}{{\tt
  astro-ph/0410281}}].

\bibitem{Bassett:2005xm}
B.~A. Bassett, S.~Tsujikawa, and D.~Wands, {\it {Inflation dynamics and
  reheating}},  {\em Rev. Mod. Phys.} {\bf 78} (2006) 537--589,
  [\href{http://arxiv.org/abs/astro-ph/0507632}{{\tt astro-ph/0507632}}].

\bibitem{Baumann:2009ds}
D.~Baumann, {\it {Inflation}},  in {\em {Theoretical Advanced Study Institute
  in Elementary Particle Physics}: {Physics of the Large and the Small}},
  pp.~523--686, 2011.
\newblock \href{http://arxiv.org/abs/0907.5424}{{\tt arXiv:0907.5424}}.

\bibitem{Sriramkumar:2009kg}
L.~Sriramkumar, {\it {An introduction to inflation and cosmological
  perturbation theory}},  \href{http://arxiv.org/abs/0904.4584}{{\tt
  arXiv:0904.4584}}.

\bibitem{Martin:2013tda}
J.~Martin, C.~Ringeval, and V.~Vennin, {\it {Encyclop\ae{}dia Inflationaris}},
  {\em Phys. Dark Univ.} {\bf 5-6} (2014) 75--235,
  [\href{http://arxiv.org/abs/1303.3787}{{\tt arXiv:1303.3787}}].

\bibitem{WMAP:2012nax}
{\bf WMAP} Collaboration, G.~Hinshaw et~al., {\it {Nine-Year Wilkinson
  Microwave Anisotropy Probe (WMAP) Observations: Cosmological Parameter
  Results}},  {\em Astrophys. J. Suppl.} {\bf 208} (2013) 19,
  [\href{http://arxiv.org/abs/1212.5226}{{\tt arXiv:1212.5226}}].

\bibitem{Planck:2018jri}
{\bf Planck} Collaboration, Y.~Akrami et~al., {\it {Planck 2018 results. X.
  Constraints on inflation}},  {\em Astron. Astrophys.} {\bf 641} (2020) A10,
  [\href{http://arxiv.org/abs/1807.06211}{{\tt arXiv:1807.06211}}].

\bibitem{Planck:2019kim}
{\bf Planck} Collaboration, Y.~Akrami et~al., {\it {Planck 2018 results. IX.
  Constraints on primordial non-Gaussianity}},  {\em Astron. Astrophys.} {\bf
  641} (2020) A9, [\href{http://arxiv.org/abs/1905.05697}{{\tt
  arXiv:1905.05697}}].

\bibitem{BICEP:2021xfz}
{\bf BICEP, Keck} Collaboration, P.~A.~R. Ade et~al., {\it {Improved
  Constraints on Primordial Gravitational Waves using Planck, WMAP, and
  BICEP/Keck Observations through the 2018 Observing Season}},  {\em Phys. Rev.
  Lett.} {\bf 127} (2021), no.~15 151301,
  [\href{http://arxiv.org/abs/2110.00483}{{\tt arXiv:2110.00483}}].

\bibitem{WMAP:2008ttx}
{\bf WMAP} Collaboration, M.~R. Nolta et~al., {\it {Five-Year Wilkinson
  Microwave Anisotropy Probe (WMAP) Observations: Angular Power Spectra}},
  {\em Astrophys. J. Suppl.} {\bf 180} (2009) 296--305,
  [\href{http://arxiv.org/abs/0803.0593}{{\tt arXiv:0803.0593}}].

\bibitem{Schwarz:2015cma}
D.~J. Schwarz, C.~J. Copi, D.~Huterer, and G.~D. Starkman, {\it {CMB Anomalies
  after Planck}},  {\em Class. Quant. Grav.} {\bf 33} (2016), no.~18 184001,
  [\href{http://arxiv.org/abs/1510.07929}{{\tt arXiv:1510.07929}}].

\bibitem{Planck:2019evm}
{\bf Planck} Collaboration, Y.~Akrami et~al., {\it {Planck 2018 results. VII.
  Isotropy and Statistics of the CMB}},  {\em Astron. Astrophys.} {\bf 641}
  (2020) A7, [\href{http://arxiv.org/abs/1906.02552}{{\tt arXiv:1906.02552}}].

\bibitem{Chluba:2015bqa}
J.~Chluba, J.~Hamann, and S.~P. Patil, {\it {Features and New Physical Scales
  in Primordial Observables: Theory and Observation}},  {\em Int. J. Mod. Phys.
  D} {\bf 24} (2015), no.~10 1530023,
  [\href{http://arxiv.org/abs/1505.01834}{{\tt arXiv:1505.01834}}].

\bibitem{Slosar:2019gvt}
A.~Slosar et~al., {\it {Scratches from the Past: Inflationary Archaeology
  through Features in the Power Spectrum of Primordial Fluctuations}},  {\em
  Bull. Am. Astron. Soc.} {\bf 51} (2019), no.~3 98,
  [\href{http://arxiv.org/abs/1903.09883}{{\tt arXiv:1903.09883}}].

\bibitem{Adams:2001vc}
J.~A. Adams, B.~Cresswell, and R.~Easther, {\it {Inflationary perturbations
  from a potential with a step}},  {\em Phys. Rev. D} {\bf 64} (2001) 123514,
  [\href{http://arxiv.org/abs/astro-ph/0102236}{{\tt astro-ph/0102236}}].

\bibitem{Hunt:2004vt}
P.~Hunt and S.~Sarkar, {\it {Multiple inflation and the WMAP 'glitches'}},
  {\em Phys. Rev. D} {\bf 70} (2004) 103518,
  [\href{http://arxiv.org/abs/astro-ph/0408138}{{\tt astro-ph/0408138}}].

\bibitem{Hunt:2007dn}
P.~Hunt and S.~Sarkar, {\it {Multiple inflation and the WMAP glitches. 2. Data
  analysis and cosmological parameter extraction}},  {\em Phys. Rev. D} {\bf
  76} (2007) 123504, [\href{http://arxiv.org/abs/0706.2443}{{\tt
  arXiv:0706.2443}}].

\bibitem{Covi:2006ci}
L.~Covi, J.~Hamann, A.~Melchiorri, A.~Slosar, and I.~Sorbera, {\it {Inflation
  and WMAP three year data: Features have a Future!}},  {\em Phys. Rev. D} {\bf
  74} (2006) 083509, [\href{http://arxiv.org/abs/astro-ph/0606452}{{\tt
  astro-ph/0606452}}].

\bibitem{Hamann:2007pa}
J.~Hamann, L.~Covi, A.~Melchiorri, and A.~Slosar, {\it {New Constraints on
  Oscillations in the Primordial Spectrum of Inflationary Perturbations}},
  {\em Phys. Rev. D} {\bf 76} (2007) 023503,
  [\href{http://arxiv.org/abs/astro-ph/0701380}{{\tt astro-ph/0701380}}].

\bibitem{Hazra:2010ve}
D.~K. Hazra, M.~Aich, R.~K. Jain, L.~Sriramkumar, and T.~Souradeep, {\it
  {Primordial features due to a step in the inflaton potential}},  {\em JCAP}
  {\bf 10} (2010) 008, [\href{http://arxiv.org/abs/1005.2175}{{\tt
  arXiv:1005.2175}}].

\bibitem{Ashoorioon:2014yua}
A.~Ashoorioon, C.~van~de Bruck, P.~Millington, and S.~Vu, {\it {Effect of
  transitions in the Planck mass during inflation on primordial power
  spectra}},  {\em Phys. Rev. D} {\bf 90} (2014) 103515,
  [\href{http://arxiv.org/abs/1406.5466}{{\tt arXiv:1406.5466}}].

\bibitem{Hunt:2015iua}
P.~Hunt and S.~Sarkar, {\it {Search for features in the spectrum of primordial
  perturbations using Planck and other datasets}},  {\em JCAP} {\bf 12} (2015)
  052, [\href{http://arxiv.org/abs/1510.03338}{{\tt arXiv:1510.03338}}].

\bibitem{Dalianis:2021iig}
I.~Dalianis, G.~P. Kodaxis, I.~D. Stamou, N.~Tetradis, and A.~Tsigkas-Kouvelis,
  {\it {Spectrum oscillations from features in the potential of single-field
  inflation}},  {\em Phys. Rev. D} {\bf 104} (2021), no.~10 103510,
  [\href{http://arxiv.org/abs/2106.02467}{{\tt arXiv:2106.02467}}].

\bibitem{Jain:2007au}
R.~K. Jain, P.~Chingangbam, and L.~Sriramkumar, {\it {On the evolution of
  tachyonic perturbations at super-Hubble scales}},  {\em JCAP} {\bf 10} (2007)
  003, [\href{http://arxiv.org/abs/astro-ph/0703762}{{\tt astro-ph/0703762}}].

\bibitem{Jain:2008dw}
R.~K. Jain, P.~Chingangbam, J.-O. Gong, L.~Sriramkumar, and T.~Souradeep, {\it
  {Punctuated inflation and the low CMB multipoles}},  {\em JCAP} {\bf 01}
  (2009) 009, [\href{http://arxiv.org/abs/0809.3915}{{\tt arXiv:0809.3915}}].

\bibitem{Jain:2009pm}
R.~K. Jain, P.~Chingangbam, L.~Sriramkumar, and T.~Souradeep, {\it {The
  tensor-to-scalar ratio in punctuated inflation}},  {\em Phys. Rev. D} {\bf
  82} (2010) 023509, [\href{http://arxiv.org/abs/0904.2518}{{\tt
  arXiv:0904.2518}}].

\bibitem{Qureshi:2016pjy}
M.~H. Qureshi, A.~Iqbal, M.~A. Malik, and T.~Souradeep, {\it {Low-$\ell$ power
  suppression in punctuated inflation}},  {\em JCAP} {\bf 04} (2017) 013,
  [\href{http://arxiv.org/abs/1610.05776}{{\tt arXiv:1610.05776}}].

\bibitem{Contaldi:2003zv}
C.~R. Contaldi, M.~Peloso, L.~Kofman, and A.~D. Linde, {\it {Suppressing the
  lower multipoles in the CMB anisotropies}},  {\em JCAP} {\bf 07} (2003) 002,
  [\href{http://arxiv.org/abs/astro-ph/0303636}{{\tt astro-ph/0303636}}].

\bibitem{Sriramkumar:2004pj}
L.~Sriramkumar and T.~Padmanabhan, {\it {Initial state of matter fields and
  trans-Planckian physics: Can CMB observations disentangle the two?}},  {\em
  Phys. Rev. D} {\bf 71} (2005) 103512,
  [\href{http://arxiv.org/abs/gr-qc/0408034}{{\tt gr-qc/0408034}}].

\bibitem{Nicholson:2007by}
G.~Nicholson and C.~R. Contaldi, {\it {The large scale CMB cut-off and the
  tensor-to-scalar ratio}},  {\em JCAP} {\bf 01} (2008) 002,
  [\href{http://arxiv.org/abs/astro-ph/0701783}{{\tt astro-ph/0701783}}].

\bibitem{Hergt:2018crm}
L.~T. Hergt, W.~J. Handley, M.~P. Hobson, and A.~N. Lasenby, {\it {Case for
  kinetically dominated initial conditions for inflation}},  {\em Phys. Rev. D}
  {\bf 100} (2019), no.~2 023502, [\href{http://arxiv.org/abs/1809.07185}{{\tt
  arXiv:1809.07185}}].

\bibitem{Ragavendra:2020old}
H.~V. Ragavendra, D.~Chowdhury, and L.~Sriramkumar, {\it {Suppression of scalar
  power on large scales and associated bispectra}},
  \href{http://arxiv.org/abs/2003.01099}{{\tt arXiv:2003.01099}}.

\bibitem{Wands:2007bd}
D.~Wands, {\it {Multiple field inflation}},  {\em Lect. Notes Phys.} {\bf 738}
  (2008) 275--304, [\href{http://arxiv.org/abs/astro-ph/0702187}{{\tt
  astro-ph/0702187}}].

\bibitem{Achucarro:2010da}
A.~Achucarro, J.-O. Gong, S.~Hardeman, G.~A. Palma, and S.~P. Patil, {\it
  {Features of heavy physics in the CMB power spectrum}},  {\em JCAP} {\bf 01}
  (2011) 030, [\href{http://arxiv.org/abs/1010.3693}{{\tt arXiv:1010.3693}}].

\bibitem{Gao:2013ota}
X.~Gao, D.~Langlois, and S.~Mizuno, {\it {Oscillatory features in the curvature
  power spectrum after a sudden turn of the inflationary trajectory}},  {\em
  JCAP} {\bf 10} (2013) 023, [\href{http://arxiv.org/abs/1306.5680}{{\tt
  arXiv:1306.5680}}].

\bibitem{Noumi:2013cfa}
T.~Noumi and M.~Yamaguchi, {\it {Primordial spectra from sudden turning
  trajectory}},  {\em JCAP} {\bf 12} (2013) 038,
  [\href{http://arxiv.org/abs/1307.7110}{{\tt arXiv:1307.7110}}].

\bibitem{Braglia:2020fms}
M.~Braglia, D.~K. Hazra, L.~Sriramkumar, and F.~Finelli, {\it {Generating
  primordial features at large scales in two field models of inflation}},  {\em
  JCAP} {\bf 08} (2020) 025, [\href{http://arxiv.org/abs/2004.00672}{{\tt
  arXiv:2004.00672}}].

\bibitem{Horndeski:1974wa}
G.~W. Horndeski, {\it {Second-order scalar-tensor field equations in a
  four-dimensional space}},  {\em Int. J. Theor. Phys.} {\bf 10} (1974)
  363--384.

\bibitem{Deffayet:2009wt}
C.~Deffayet, G.~Esposito-Farese, and A.~Vikman, {\it {Covariant Galileon}},
  {\em Phys. Rev. D} {\bf 79} (2009) 084003,
  [\href{http://arxiv.org/abs/0901.1314}{{\tt arXiv:0901.1314}}].

\bibitem{Deffayet:2009mn}
C.~Deffayet, S.~Deser, and G.~Esposito-Farese, {\it {Generalized Galileons: All
  scalar models whose curved background extensions maintain second-order field
  equations and stress-tensors}},  {\em Phys. Rev. D} {\bf 80} (2009) 064015,
  [\href{http://arxiv.org/abs/0906.1967}{{\tt arXiv:0906.1967}}].

\bibitem{Kobayashi:2011nu}
T.~Kobayashi, M.~Yamaguchi, and J.~Yokoyama, {\it {Generalized G-inflation:
  Inflation with the most general second-order field equations}},  {\em Prog.
  Theor. Phys.} {\bf 126} (2011) 511--529,
  [\href{http://arxiv.org/abs/1105.5723}{{\tt arXiv:1105.5723}}].

\bibitem{Karydas:2021wmx}
S.~Karydas, E.~Papantonopoulos, and E.~N. Saridakis, {\it {Successful Higgs
  inflation from combined nonminimal and derivative couplings}},  {\em Phys.
  Rev. D} {\bf 104} (2021), no.~2 023530,
  [\href{http://arxiv.org/abs/2102.08450}{{\tt arXiv:2102.08450}}].

\bibitem{Amendola:1993uh}
L.~Amendola, {\it {Cosmology with nonminimal derivative couplings}},  {\em
  Phys. Lett. B} {\bf 301} (1993) 175--182,
  [\href{http://arxiv.org/abs/gr-qc/9302010}{{\tt gr-qc/9302010}}].

\bibitem{Tumurtushaa:2019bmc}
G.~Tumurtushaa, {\it {Inflation with Derivative Self-interaction and Coupling
  to Gravity}},  {\em Eur. Phys. J. C} {\bf 79} (2019), no.~11 920,
  [\href{http://arxiv.org/abs/1903.05354}{{\tt arXiv:1903.05354}}].

\bibitem{Rinaldi:2012vy}
M.~Rinaldi, {\it {Black holes with non-minimal derivative coupling}},  {\em
  Phys. Rev. D} {\bf 86} (2012) 084048,
  [\href{http://arxiv.org/abs/1208.0103}{{\tt arXiv:1208.0103}}].

\bibitem{Anabalon:2013oea}
A.~Anabalon, A.~Cisterna, and J.~Oliva, {\it {Asymptotically locally AdS and
  flat black holes in Horndeski theory}},  {\em Phys. Rev. D} {\bf 89} (2014)
  084050, [\href{http://arxiv.org/abs/1312.3597}{{\tt arXiv:1312.3597}}].

\bibitem{Ema:2015oaa}
Y.~Ema, R.~Jinno, K.~Mukaida, and K.~Nakayama, {\it {Particle Production after
  Inflation with Non-minimal Derivative Coupling to Gravity}},  {\em JCAP} {\bf
  10} (2015) 020, [\href{http://arxiv.org/abs/1504.07119}{{\tt
  arXiv:1504.07119}}].

\bibitem{Ghalee:2013ada}
A.~Ghalee, {\it {A new phase of scalar field with a kinetic term non-minimally
  coupled to gravity}},  {\em Phys. Lett. B} {\bf 724} (2013) 198--202,
  [\href{http://arxiv.org/abs/1303.0532}{{\tt arXiv:1303.0532}}].

\bibitem{Fu:2019ttf}
C.~Fu, P.~Wu, and H.~Yu, {\it {Primordial Black Holes from Inflation with
  Nonminimal Derivative Coupling}},  {\em Phys. Rev. D} {\bf 100} (2019), no.~6
  063532, [\href{http://arxiv.org/abs/1907.05042}{{\tt arXiv:1907.05042}}].

\bibitem{Heydari:2021gea}
S.~Heydari and K.~Karami, {\it {Primordial black holes in nonminimal derivative
  coupling inflation with quartic potential and reheating consideration}},
  {\em Eur. Phys. J. C} {\bf 82} (2022), no.~1 83,
  [\href{http://arxiv.org/abs/2107.10550}{{\tt arXiv:2107.10550}}].

\bibitem{Heydari:2021qsr}
S.~Heydari and K.~Karami, {\it {Primordial black holes ensued from exponential
  potential and coupling parameter in nonminimal derivative inflation model}},
  {\em JCAP} {\bf 03} (2022), no.~03 033,
  [\href{http://arxiv.org/abs/2111.00494}{{\tt arXiv:2111.00494}}].

\bibitem{Lewis:2002ah}
A.~Lewis and S.~Bridle, {\it {Cosmological parameters from CMB and other data:
  A Monte Carlo approach}},  {\em Phys. Rev. D} {\bf 66} (2002) 103511,
  [\href{http://arxiv.org/abs/astro-ph/0205436}{{\tt astro-ph/0205436}}].

\bibitem{Kobayashi:2019hrl}
T.~Kobayashi, {\it {Horndeski theory and beyond: a review}},  {\em Rept. Prog.
  Phys.} {\bf 82} (2019), no.~8 086901,
  [\href{http://arxiv.org/abs/1901.07183}{{\tt arXiv:1901.07183}}].

\bibitem{Quiros:2017gsu}
I.~Quiros, T.~Gonzalez, U.~Nucamendi, R.~Garc\'\i{}a-Salcedo, F.~A.
  Horta-Rangel, and J.~Saavedra, {\it {On the phantom barrier crossing and the
  bounds on the speed of sound in non-minimal derivative coupling theories}},
  {\em Class. Quant. Grav.} {\bf 35} (2018), no.~7 075005,
  [\href{http://arxiv.org/abs/1707.03885}{{\tt arXiv:1707.03885}}].

\bibitem{Hsu:2004vr}
S.~D.~H. Hsu, A.~Jenkins, and M.~B. Wise, {\it {Gradient instability for w
  \ensuremath{<} -1}},  {\em Phys. Lett. B} {\bf 597} (2004) 270--274,
  [\href{http://arxiv.org/abs/astro-ph/0406043}{{\tt astro-ph/0406043}}].

\bibitem{LIGOScientific:2017zic}
{\bf LIGO Scientific, Virgo, Fermi-GBM, INTEGRAL} Collaboration, B.~P. Abbott
  et~al., {\it {Gravitational Waves and Gamma-rays from a Binary Neutron Star
  Merger: GW170817 and GRB 170817A}},  {\em Astrophys. J. Lett.} {\bf 848}
  (2017), no.~2 L13, [\href{http://arxiv.org/abs/1710.05834}{{\tt
  arXiv:1710.05834}}].

\bibitem{Gong:2017kim}
Y.~Gong, E.~Papantonopoulos, and Z.~Yi, {\it {Constraints on
  scalar\textendash{}tensor theory of gravity by the recent observational
  results on gravitational waves}},  {\em Eur. Phys. J. C} {\bf 78} (2018),
  no.~9 738, [\href{http://arxiv.org/abs/1711.04102}{{\tt arXiv:1711.04102}}].

\bibitem{Dalianis:2019vit}
I.~Dalianis, S.~Karydas, and E.~Papantonopoulos, {\it {Generalized Non-Minimal
  Derivative Coupling: Application to Inflation and Primordial Black Hole
  Production}},  {\em JCAP} {\bf 06} (2020) 040,
  [\href{http://arxiv.org/abs/1910.00622}{{\tt arXiv:1910.00622}}].

\bibitem{Braglia:2021rej}
M.~Braglia, X.~Chen, and D.~K. Hazra, {\it {Primordial Standard Clock Models
  and CMB Residual Anomalies}},  \href{http://arxiv.org/abs/2108.10110}{{\tt
  arXiv:2108.10110}}.

\bibitem{Hazra:2021eqk}
D.~K. Hazra, D.~Paoletti, I.~Debono, A.~Shafieloo, G.~F. Smoot, and A.~A.
  Starobinsky, {\it {Inflation story: slow-roll and beyond}},  {\em JCAP} {\bf
  12} (2021), no.~12 038, [\href{http://arxiv.org/abs/2107.09460}{{\tt
  arXiv:2107.09460}}].

\bibitem{Braglia:2021sun}
M.~Braglia, X.~Chen, and D.~K. Hazra, {\it {Uncovering the History of Cosmic
  Inflation from Anomalies in Cosmic Microwave Background Spectra}},
  \href{http://arxiv.org/abs/2106.07546}{{\tt arXiv:2106.07546}}.

\bibitem{Hazra:2014jka}
D.~K. Hazra, A.~Shafieloo, G.~F. Smoot, and A.~A. Starobinsky, {\it {Inflation
  with Whip-Shaped Suppressed Scalar Power Spectra}},  {\em Phys. Rev. Lett.}
  {\bf 113} (2014), no.~7 071301, [\href{http://arxiv.org/abs/1404.0360}{{\tt
  arXiv:1404.0360}}].

\bibitem{Planck:2019nip}
{\bf Planck} Collaboration, N.~Aghanim et~al., {\it {Planck 2018 results. V.
  CMB power spectra and likelihoods}},  {\em Astron. Astrophys.} {\bf 641}
  (2020) A5, [\href{http://arxiv.org/abs/1907.12875}{{\tt arXiv:1907.12875}}].

\bibitem{Lewis:1999bs}
A.~Lewis, A.~Challinor, and A.~Lasenby, {\it {Efficient computation of CMB
  anisotropies in closed FRW models}},  {\em Astrophys. J.} {\bf 538} (2000)
  473--476, [\href{http://arxiv.org/abs/astro-ph/9911177}{{\tt
  astro-ph/9911177}}].

\bibitem{Powel:2009bob}
M.~Powell, {\it {``The BOBYQA Algorithm for Bound Constrained Optimization
  without Derivatives''}},  {\em Technical Report, Department of Applied
  Mathematics and Theoretical Physics} (2009).

\bibitem{Lewis:2019xzd}
A.~Lewis, {\it {GetDist: a Python package for analysing Monte Carlo samples}},
  \href{http://arxiv.org/abs/1910.13970}{{\tt arXiv:1910.13970}}.

\bibitem{Gelman:1992zz}
A.~Gelman and D.~B. Rubin, {\it {Inference from Iterative Simulation Using
  Multiple Sequences}},  {\em Statist. Sci.} {\bf 7} (1992) 457--472.

\bibitem{Braglia:2021ckn}
M.~Braglia, X.~Chen, and D.~K. Hazra, {\it {Comparing multi-field primordial
  feature models with the Planck data}},  {\em JCAP} {\bf 06} (2021) 005,
  [\href{http://arxiv.org/abs/2103.03025}{{\tt arXiv:2103.03025}}].

\bibitem{Hazra:2017joc}
D.~K. Hazra, D.~Paoletti, M.~Ballardini, F.~Finelli, A.~Shafieloo, G.~F. Smoot,
  and A.~A. Starobinsky, {\it {Probing features in inflaton potential and
  reionization history with future CMB space observations}},  {\em JCAP} {\bf
  02} (2018) 017, [\href{http://arxiv.org/abs/1710.01205}{{\tt
  arXiv:1710.01205}}].

\bibitem{Domenech:2019cyh}
G.~Dom\`enech and M.~Kamionkowski, {\it {Lensing anomaly and oscillations in
  the primordial power spectrum}},  {\em JCAP} {\bf 11} (2019) 040,
  [\href{http://arxiv.org/abs/1905.04323}{{\tt arXiv:1905.04323}}].

\bibitem{Domenech:2020qay}
G.~Dom\`enech, X.~Chen, M.~Kamionkowski, and A.~Loeb, {\it {Planck residuals
  anomaly as a fingerprint of alternative scenarios to inflation}},  {\em JCAP}
  {\bf 10} (2020) 005, [\href{http://arxiv.org/abs/2005.08998}{{\tt
  arXiv:2005.08998}}].

\bibitem{Chabanier:2019eai}
S.~Chabanier, M.~Millea, and N.~Palanque-Delabrouille, {\it {Matter power
  spectrum: from Ly$\alpha$ forest to CMB scales}},  {\em Mon. Not. Roy.
  Astron. Soc.} {\bf 489} (2019), no.~2 2247--2253,
  [\href{http://arxiv.org/abs/1905.08103}{{\tt arXiv:1905.08103}}].

\bibitem{Chen:2016vvw}
X.~Chen, C.~Dvorkin, Z.~Huang, M.~H. Namjoo, and L.~Verde, {\it {The Future of
  Primordial Features with Large-Scale Structure Surveys}},  {\em JCAP} {\bf
  11} (2016) 014, [\href{http://arxiv.org/abs/1605.09365}{{\tt
  arXiv:1605.09365}}].

\bibitem{Ballardini:2016hpi}
M.~Ballardini, F.~Finelli, C.~Fedeli, and L.~Moscardini, {\it {Probing
  primordial features with future galaxy surveys}},  {\em JCAP} {\bf 10} (2016)
  041, [\href{http://arxiv.org/abs/1606.03747}{{\tt arXiv:1606.03747}}].
  [Erratum: JCAP 04, E01 (2018)].

\bibitem{Ballardini:2017qwq}
M.~Ballardini, F.~Finelli, R.~Maartens, and L.~Moscardini, {\it {Probing
  primordial features with next-generation photometric and radio surveys}},
  {\em JCAP} {\bf 04} (2018) 044, [\href{http://arxiv.org/abs/1712.07425}{{\tt
  arXiv:1712.07425}}].

\bibitem{Palma:2017wxu}
G.~A. Palma, D.~Sapone, and S.~Sypsas, {\it {Constraints on inflation with LSS
  surveys: features in the primordial power spectrum}},  {\em JCAP} {\bf 06}
  (2018) 004, [\href{http://arxiv.org/abs/1710.02570}{{\tt arXiv:1710.02570}}].

\bibitem{LHuillier:2017lgm}
B.~L'Huillier, A.~Shafieloo, D.~K. Hazra, G.~F. Smoot, and A.~A. Starobinsky,
  {\it {Probing features in the primordial perturbation spectrum with
  large-scale structure data}},  {\em Mon. Not. Roy. Astron. Soc.} {\bf 477}
  (2018), no.~2 2503--2512, [\href{http://arxiv.org/abs/1710.10987}{{\tt
  arXiv:1710.10987}}].

\bibitem{Beutler:2019ojk}
F.~Beutler, M.~Biagetti, D.~Green, A.~Slosar, and B.~Wallisch, {\it {Primordial
  Features from Linear to Nonlinear Scales}},  {\em Phys. Rev. Res.} {\bf 1}
  (2019), no.~3 033209, [\href{http://arxiv.org/abs/1906.08758}{{\tt
  arXiv:1906.08758}}].

\bibitem{Ballardini:2019tuc}
M.~Ballardini, R.~Murgia, M.~Baldi, F.~Finelli, and M.~Viel, {\it {Non-linear
  damping of superimposed primordial oscillations on the matter power spectrum
  in galaxy surveys}},  {\em JCAP} {\bf 04} (2020), no.~04 030,
  [\href{http://arxiv.org/abs/1912.12499}{{\tt arXiv:1912.12499}}].

\bibitem{Esmaeilian:2020rbq}
M.~S. Esmaeilian, M.~Farhang, and S.~Khodabakhshi, {\it {Detectable
  data--driven features in the primordial scalar power spectrum}},  {\em
  Astrophys. J.} {\bf 912} (2021), no.~2 104,
  [\href{http://arxiv.org/abs/2011.14774}{{\tt arXiv:2011.14774}}].

\bibitem{Li:2021jvz}
Y.~Li, H.-M. Zhu, and B.~Li, {\it {Nonlinear reconstruction of features in the
  primordial power spectrum from large-scale structure}},
  \href{http://arxiv.org/abs/2102.09007}{{\tt arXiv:2102.09007}}.

\bibitem{Ballardini:2022wzu}
M.~Ballardini, F.~Finelli, F.~Marulli, L.~Moscardini, and A.~Veropalumbo, {\it
  {New constraints on primordial features from the galaxy two-point correlation
  function}},  \href{http://arxiv.org/abs/2202.08819}{{\tt arXiv:2202.08819}}.

\bibitem{Debono:2020emh}
I.~Debono, D.~K. Hazra, A.~Shafieloo, G.~F. Smoot, and A.~A. Starobinsky, {\it
  {Constraints on features in the inflationary potential from future Euclid
  data}},  {\em Mon. Not. Roy. Astron. Soc.} {\bf 496} (2020), no.~3
  3448--3468, [\href{http://arxiv.org/abs/2003.05262}{{\tt arXiv:2003.05262}}].

\bibitem{Chandra:2022utq}
D.~Chandra and S.~Pal, {\it {Investigating the Constraints on Primordial
  Features with Future Cosmic Microwave Background and Galaxy Surveys}},
  \href{http://arxiv.org/abs/2205.01164}{{\tt arXiv:2205.01164}}.

\bibitem{Chen:2016zuu}
X.~Chen, P.~D. Meerburg, and M.~M\"unchmeyer, {\it {The Future of Primordial
  Features with 21 cm Tomography}},  {\em JCAP} {\bf 09} (2016) 023,
  [\href{http://arxiv.org/abs/1605.09364}{{\tt arXiv:1605.09364}}].

\bibitem{Xu:2016kwz}
Y.~Xu, J.~Hamann, and X.~Chen, {\it {Precise measurements of inflationary
  features with 21 cm observations}},  {\em Phys. Rev. D} {\bf 94} (2016),
  no.~12 123518, [\href{http://arxiv.org/abs/1607.00817}{{\tt
  arXiv:1607.00817}}].

\bibitem{Balaji:2022zur}
S.~Balaji, H.~V. Ragavendra, S.~K. Sethi, J.~Silk, and L.~Sriramkumar, {\it
  {Observing nulling of primordial correlations via the 21 cm signal}},
  \href{http://arxiv.org/abs/2206.06386}{{\tt arXiv:2206.06386}}.

\bibitem{Kogut:2011xw}
A.~Kogut et~al., {\it {The Primordial Inflation Explorer (PIXIE): A Nulling
  Polarimeter for Cosmic Microwave Background Observations}},  {\em JCAP} {\bf
  07} (2011) 025, [\href{http://arxiv.org/abs/1105.2044}{{\tt
  arXiv:1105.2044}}].

\bibitem{PRISM:2013fvg}
{\bf PRISM} Collaboration, P.~Andr\'e et~al., {\it {PRISM (Polarized Radiation
  Imaging and Spectroscopy Mission): An Extended White Paper}},  {\em JCAP}
  {\bf 02} (2014) 006, [\href{http://arxiv.org/abs/1310.1554}{{\tt
  arXiv:1310.1554}}].

\bibitem{Chen:2006xjb}
X.~Chen, R.~Easther, and E.~A. Lim, {\it {Large Non-Gaussianities in Single
  Field Inflation}},  {\em JCAP} {\bf 06} (2007) 023,
  [\href{http://arxiv.org/abs/astro-ph/0611645}{{\tt astro-ph/0611645}}].

\bibitem{Chen:2008wn}
X.~Chen, R.~Easther, and E.~A. Lim, {\it {Generation and Characterization of
  Large Non-Gaussianities in Single Field Inflation}},  {\em JCAP} {\bf 04}
  (2008) 010, [\href{http://arxiv.org/abs/0801.3295}{{\tt arXiv:0801.3295}}].

\bibitem{Adshead:2011jq}
P.~Adshead, C.~Dvorkin, W.~Hu, and E.~A. Lim, {\it {Non-Gaussianity from Step
  Features in the Inflationary Potential}},  {\em Phys. Rev. D} {\bf 85} (2012)
  023531, [\href{http://arxiv.org/abs/1110.3050}{{\tt arXiv:1110.3050}}].

\bibitem{Hazra:2012yn}
D.~K. Hazra, L.~Sriramkumar, and J.~Martin, {\it {BINGO: A code for the
  efficient computation of the scalar bi-spectrum}},  {\em JCAP} {\bf 05}
  (2013) 026, [\href{http://arxiv.org/abs/1201.0926}{{\tt arXiv:1201.0926}}].

\bibitem{Fergusson:2014hya}
J.~R. Fergusson, H.~F. Gruetjen, E.~P.~S. Shellard, and M.~Liguori, {\it
  {Combining power spectrum and bispectrum measurements to detect oscillatory
  features}},  {\em Phys. Rev. D} {\bf 91} (2015), no.~2 023502,
  [\href{http://arxiv.org/abs/1410.5114}{{\tt arXiv:1410.5114}}].

\bibitem{Fergusson:2014tza}
J.~R. Fergusson, H.~F. Gruetjen, E.~P.~S. Shellard, and B.~Wallisch, {\it
  {Polyspectra searches for sharp oscillatory features in cosmic microwave sky
  data}},  {\em Phys. Rev. D} {\bf 91} (2015), no.~12 123506,
  [\href{http://arxiv.org/abs/1412.6152}{{\tt arXiv:1412.6152}}].

\bibitem{CMB-S4:2016ple}
{\bf CMB-S4} Collaboration, K.~N. Abazajian et~al., {\it {CMB-S4 Science Book,
  First Edition}},  \href{http://arxiv.org/abs/1610.02743}{{\tt
  arXiv:1610.02743}}.

\bibitem{LiteBIRD:2020khw}
{\bf LiteBIRD} Collaboration, M.~Hazumi et~al., {\it {LiteBIRD: JAXA's new
  strategic L-class mission for all-sky surveys of cosmic microwave background
  polarization}},  {\em Proc. SPIE Int. Soc. Opt. Eng.} {\bf 11443} (2020)
  114432F, [\href{http://arxiv.org/abs/2101.12449}{{\tt arXiv:2101.12449}}].

\bibitem{2019BAAS51g.286L}
A.~Lee et~al., {\it {LiteBIRD: an all-sky cosmic microwave background probe of
  inflation}},  {\em Bulletin of the American Astronomical Society} {\bf 51}
  (2019) 286.

\bibitem{SimonsObservatory:2018koc}
{\bf Simons Observatory} Collaboration, P.~Ade et~al., {\it {The Simons
  Observatory: Science goals and forecasts}},  {\em JCAP} {\bf 02} (2019) 056,
  [\href{http://arxiv.org/abs/1808.07445}{{\tt arXiv:1808.07445}}].

\bibitem{SimonsObservatory:2019qwx}
{\bf Simons Observatory} Collaboration, M.~H. Abitbol et~al., {\it {The Simons
  Observatory: Astro2020 Decadal Project Whitepaper}},  {\em Bull. Am. Astron.
  Soc.} {\bf 51} (2019) 147, [\href{http://arxiv.org/abs/1907.08284}{{\tt
  arXiv:1907.08284}}].

\bibitem{cmb-bharat}
{CMB-Bharat}, {\it \url{http://cmb-bharat.in}}, .

\end{thebibliography}\endgroup


\providecommand{\href}[2]{#2}\begingroup\raggedright\endgroup
\end{document}